\documentclass[useAMS,usenatbib,pdftex,usegraphicx]{mn2e}
\usepackage{pdflscape}

\newcommand{\Lnu}{\ifmmode L_{\rm \nu} \else $L_{\rm \nu}$\fi}
\newcommand{\Llambda}{\ifmmode L_{\rm \lambda} \else $L_{\rm \lambda}$\fi}
\newcommand{\LA}{\ifmmode L_{\rm 5100} \else $L_{\rm 5100}$\fi}
\newcommand{\Lg}{\ifmmode L_{\rm g(r)} \else $L_{\rm g(r)}$\fi}
\newcommand{\Ln}{\ifmmode L_{\rm n(r)} \else $L_{\rm n(r)}$\fi}
\newcommand{\Lo}{\ifmmode L_{\rm 0} \else $L_{\rm 0}$\fi}
\newcommand{\Ro}{\ifmmode R_{\rm 0} \else $R_{\rm 0}$\fi}
\newcommand{\Rhalf}{\ifmmode R_{\rm 1/2} \else $R_{\rm 1/2}$\fi}
\newcommand{\Rg}{\ifmmode R_{\rm g} \else $R_{\rm g}$\fi}
\newcommand{\Rms}{\ifmmode R_{\rm ms} \else $R_{\rm ms}$\fi}
\newcommand{\Rin}{\ifmmode R_{\rm in} \else $R_{\rm in}$\fi}
\newcommand{\Rout}{\ifmmode R_{\rm out} \else $R_{\rm out}$\fi}
\newcommand{\RISCO}{\ifmmode R_{\rm in} \else $R_{\rm in}$\fi}
\newcommand{\ro}{\ifmmode r_{\rm 0} \else $r_{\rm 0}$\fi}
\newcommand{\rhalf}{\ifmmode r_{\rm 1/2} \else $r_{\rm 1/2}$\fi}
\newcommand{\rg}{\ifmmode r_{\rm g} \else $r_{\rm g}$\fi}
\newcommand{\rms}{\ifmmode r_{\rm ms} \else $r_{\rm ms}$\fi}
\newcommand{\rin}{\ifmmode r_{\rm in} \else $r_{\rm in}$\fi}
\newcommand{\rout}{\ifmmode r_{\rm out} \else $r_{\rm out}$\fi}
\newcommand{\rISCO}{\ifmmode r_{\rm in} \else $r_{\rm in}$\fi}
\newcommand{\Fnu}{\ifmmode F_{\nu} \else $F_{\nu}$\fi}
\newcommand{\Flambda}{\ifmmode F_{\lambda} \else $F_{\lambda}$\fi}
\newcommand{\Mdot}{\ifmmode \dot{M} \else $\dot{M}$\fi}
\newcommand{\mdot}{\ifmmode \dot{m} \else $\dot{m}$\fi}
\newcommand{\Mrdot}{\ifmmode \dot{M}\left(r \right) \else $\dot{M}\left(r \right)$\fi}
\newcommand{\MRdot}{\ifmmode \dot{M}\left(R \right) \else $\dot{M}\left(R \right)$\fi}
\newcommand{\Mrindot}{\ifmmode \dot{M}\left(r_{ISCO} \right) \else $\dot{M}\left(r_{ISCO} \right)$\fi}
\newcommand{\Mroutdot}{\ifmmode \dot{M}\left(r_{out} \right) \else $\dot{M}\left(r_{out} \right)$\fi}
\newcommand{\MBHdot}{\ifmmode \dot{M}_{BH} \else $\dot{M}_{BH}$\fi}
\newcommand{\MBHdotexpct}{\ifmmode \dot{M}_{BHexpected} \else $\dot{M}_{BHexpected}$\fi}
\newcommand{\Medddot}{\ifmmode \dot{M}_{edd} \else $\dot{M}_{edd}$\fi}
\newcommand{\Moutdot}{\ifmmode \dot{M}_{out} \else $\dot{M}_{out}$\fi}
\newcommand{\Mindot}{\ifmmode \dot{M}_{in} \else $\dot{M}_{in}$\fi}
\newcommand{\Mwinddot}{\ifmmode \dot{M}_{wind} \else $\dot{M}_{wind}$\fi}
\newcommand{\MBH}{\ifmmode M_{\rm BH} \else $M_{\rm BH}$\fi}
\newcommand{\mbh}{\ifmmode M_{\rm BH} \else $M_{\rm BH}$\fi}
\newcommand{\Mexp}{\ifmmode M_{\rm 8} \else $M_{\rm 8}$\fi}

\newcommand{\Msun}{\ifmmode M_{\odot} \else $M_{\odot}$\fi}
\newcommand{\msun}{\ifmmode M_{\odot} \else $M_{\odot}$\fi}
\newcommand{\Msunyr}{\ifmmode M_{\odot}/yr \else $M_{\odot}/yr$\fi}
\newcommand{\msunyr}{\ifmmode M_{\odot}/yr \else $M_{\odot}/yr$\fi}
\newcommand{\avisc}{\ifmmode \alpha_{visc} \else $\alpha_{visc}$\fi}

\newcommand{\Halpha}{\ifmmode {\rm H}\alpha \else H$\alpha$\fi}
\newcommand{\Hbeta}{\ifmmode {\rm H}\beta \else H$\beta$\fi}
\newcommand{\hb}{\ifmmode {\rm H}\beta \else H$\beta$\fi}
\newcommand{\MgII}{\ifmmode {\rm Mg}\,\textsc{ii}\,\lambda2798 \else Mg\,{\sc ii}\,$\lambda2798$\fi}
\newcommand{\mgii}{\ifmmode {\rm Mg}{\textsc{ii}} \else Mg\,{\sc ii}\fi}
\newcommand{\CIV}{\ifmmode {\rm C}\,\textsc{iv}\,\lambda1549 \else C\,{\sc iv}\,$\lambda1549$\fi}
\newcommand{\civ}{\ifmmode {\rm C}\,\textsc{iv} \else C\,{\sc iv}\fi}

\newcommand{\oi}{\ifmmode \left[{\rm O}\,\textsc{i}\right] \else [O\,{\sc i}]\fi}
\newcommand{\OI}{\ifmmode \left[{\rm O}\,\textsc{i}\right]\,\lambda6300 \else [O\,{\sc i}]$\,\lambda6300$ \fi}

\newcommand{\oii}{\ifmmode \left[{\rm O}\,\textsc{ii}\right] \else [O\,{\sc ii}]\fi}
\newcommand{\OII}{\ifmmode \left[{\rm O}\,\textsc{ii}\right]\,\lambda3727 \else [O\,{\sc ii}]\,$\lambda3727$ \fi}
\newcommand{\oiii}{\ifmmode \left[{\rm O}\,\textsc{iii}\right] \else [O\,{\sc iii}]\fi}
\newcommand{\OIII}{\ifmmode \left[{\rm O}\,\textsc{iii}\right]\,\lambda5007 \else [O\,{\sc iii}]\,$\lambda5007$\fi}

\newcommand{\ld}{\ifmmode {\rm lt-days} \else lt-days \fi}

\newcommand{\ergs}{\ifmmode {\rm erg\,s}^{-1} \else erg\,s$^{-1}$ \fi}
\newcommand{\ergcms}{\ifmmode {\rm erg\,cm}^{-2}\,{\rm s}^{-1} \else erg\,cm$^{-2}$\,s$^{-1}$\fi}
\newcommand{\ergcmsA}{\ifmmode{\rm erg}\, {\rm cm}^{-2}\,{\rm s}^{-1}\,{\rm\AA}^{-1} \else erg\, cm$^{-2}$\, s$^{-1}$\, \AA$^{-1}$\fi}
\newcommand{\ergcmsHz}{\ifmmode{\rm erg\,cm}^{-2}\,{\rm s}^{-1}\,{\rm Hz}^{-1} \else erg\,cm$^{-2}$\,s$^{-1}$\,Hz$^{-1}$\fi}
\newcommand{\phcms}{\ifmmode {\rm ph\,cm}^{-2}\,{\rm s}^{-1} \else ,ph\,cm$^{-2}$\,s$^{-1}$\fi}
\newcommand{\phcmsA}{\ifmmode {\rm ph\,cm}^{-2}\,{\rm s}^{-1}\,{\rm\AA}^{-1} \else ph\,cm$^{-2}$\,s$^{-1}$\,\AA$^{-1}$\fi}

\newcommand{\Lsun}{\ifmmode L_{\odot} \else $L_{\odot}$\fi}
\newcommand{\auvo}{\ifmmode \alpha_{\nu,{\rm UVO}} \else $\alpha_{\nu,{\rm UVO}}$\fi}
\newcommand{\Luv}{\ifmmode L_{1450} \else $L_{1450}$\fi}
\newcommand{\Lop}{\ifmmode L_{5100} \else $L_{5100}$\fi}
\newcommand{\Lthree}{\ifmmode L_{3000} \else $L_{3000}$\fi}
\newcommand{\lLthree}{\ifmmode \log\left(\Lthree/\ergs\right) \else $\log\left(\Lthree/\ergs\right)$\fi}
\newcommand{\lledd}{\ifmmode L/L_{\rm Edd} \else $L/L_{\rm Edd}$\fi}
\newcommand{\Ledd}{\ifmmode L/L_{\rm Edd} \else $L/L_{\rm Edd}$\fi}
\newcommand{\lamLlam}{\ifmmode \lambda L_{\lambda} \else $\lambda L_{\lambda}$\fi}
\newcommand{\Lbol}{\ifmmode L_{\rm bol} \else $L_{\rm bol}$\fi}
\newcommand{\lLbol}{\ifmmode \log\left(\Lbol/\ergs\right) \else $\log\left(\Lbol/\ergs\right)$\fi}

\newcommand{\Fthree}{\ifmmode F_{3000} \else $F_{3000}$\fi}

\title[Accretion Discs]
{Active galactic nuclei at $z\sim 1.5$: I. Spectral energy distribution and accretion discs}
\author[D. M. Capellupo et al.]{D. M. Capellupo$^{1}$
\thanks{E-mail:danielc@wise.tau.ac.il (DMC)},
H. Netzer$^{1}$, P. Lira$^{2}$, B. Trakhtenbrot$^{3}$
\thanks{Zwicky Postdoctoral Fellow},
Juli\'{a}n Mej\'{i}a-Restrepo$^{2}$\\
$^{1}$School of Physics and Astronomy, Tel Aviv University, Tel Aviv 69978, Israel\\
$^{2}$Departamento de Astronom\'{i}a, Universidad de Chile, Camino del Observatorio 1515, Santiago, Chile\\
$^{3}$Institute for Astronomy, Dept. of Physics, ETH Zurich, Wolfgang-Pauli-Strasse 27, CH-8093 Zurich, Switzerland}

\begin{document}


\pagerange{\pageref{firstpage}--\pageref{lastpage}} \pubyear{2002}

\maketitle

\label{firstpage}

\begin{abstract}
The physics of active super massive black holes (BHs) is governed by their mass (\MBH), spin ($a_*$) and accretion rate (\Mdot). This work is the first in a series of papers with the aim of testing how these parameters determine the observable attributes of active galactic nuclei (AGN). We have selected a sample in a narrow redshift range, centered on z $\sim$ 1.55, that covers a wide range in $M_{BH}$ and $\dot{M}$, and are observing them with {\it X-shooter}, covering rest wavelengths $\sim$1200--9800 \AA. The current work covers 30 such objects and focuses on the origin of the AGN spectral energy distribution (SED). After estimating \mbh\ and \Mdot\ based on each observed SED, we use thin AD models and a Bayesian analysis to fit the observed SEDs in our sample.
We are able to fit 22/30 of the SEDs. Out of the remaining 8 SEDs, 3 can be fit by the thin AD model by correcting the observed SED for reddening within the host galaxy and 4 can be fit by adding a disc wind to the model. In four of these 8 sources, Milky Way-type extinction, with the strong 2175\AA\ feature, provides the best reddening correction.
The distribution in spin parameter covers the entire range, from $-1$ to 0.998, and the most massive BHs have spin parameters greater than 0.7. This is consistent with the ``spin-up'' model of BH evolution. Altogether, these results indicate that thin ADs are indeed the main power houses of AGN, and earlier claims to the contrary are likely affected by variability and a limited observed wavelength range.
\end{abstract}

\begin{keywords}
galaxies: active -- quasars:general -- 
\end{keywords}

\section{Introduction}
\label{intro}

The physics of active super massive black holes (SMBHs) is governed by three key parameters: their mass (\MBH), spin ($a_*$) and accretion rate (\Mdot). 
To test how these parameters determine the observable attributes of active galactic nuclei (AGN), one needs to be able to study, in detail, a large number of sources 
covering the widest possible range in these properties. This is not a simple task given the redshift evolution and intrinsic variability of such sources, as well as the limited energy range provided by most instruments.
{\it X-shooter} \citep{Vernet11}, at the VLT, offers a way to tackle these issues. Its
high sensitivity allows one to select an AGN sample that covers a wide range of properties, and the ability to cover, simultaneously, a very wide wavelength range (3100\AA\ to the K-band) helps solve the problem of time variation.

This paper presents {\it X-shooter} observations of a unique sample of AGN at $z \simeq 1.55$, selected by both their BH mass and Eddington ratio, \Ledd. The redshift was chosen to
allow simultaneous observations of the four strong emission lines that are commonly used to measure BH mass via the ``single epoch mass determination'' method,
which uses reverberation mapping (RM) based correlations between the continuum luminosity at certain wavelengths and time lags of various broad emission lines 
\cite[e.g.][and references therein]{Kaspi00,Kaspi05,Bentz09,Bentz13}. 
The wide wavelength band and resolution of {\it X-shooter} also allows us to detect other, weaker broad and narrow emission lines, representing a large range in ionization, excitation, and critical density.
We select our AGN sample in order to evenly cover as much of the known $M_{BH}$ and $L/L_{Edd}$ parameter space as possible (see Section 2 for more details on the sample selection).
Unfortunately, little is known about $a_*$, except for a few AGN at much lower redshift \citep[and references therein]{Brenneman13,Reynolds13}.

There are three important areas of AGN science that we intend to address with this unique
sample, and each will be explored in a different paper.
The first issue, and the topic of the current work, is the origin of the AGN spectral energy distribution (SED).
The second aim is to explore the physics of the broad emission lines (BELs) in type-I AGN spectra as a function of $M_{BH}$ or $L/L_{Edd}$,
in a way which is independent of line and continuum variations.
Finally, our sample provides a unique benchmark to compare $M_{BH}$ estimates and their dependency on the emission-line 
profiles of H$\alpha$, H$\beta$, MgII 2800\AA, and CIV 1549\AA. With single-epoch spectra that 
cover all of these emission lines, over a large range in BH mass and \Ledd, we can identify the most reliable methods and determine whether the mass determinations depend on the accretion disc and/or BH properties.

As mentioned above, the current work focuses on the origin of the AGN SED, with special
emphasis on models of geometrically thin, optically thick accretion discs (ADs). Such
models follow the general ideas presented in \citet[][hereafter SS73]{Shakura73} and 
include various improvements like general relativistic (GR) corrections, radiative transfer in the disc atmosphere, and disc winds \citep[e.g.][and references therein]{Hubeny01,Davis11,Slone12}. There are several ``standard'' models of this type in the literature, which makes a comparison with observed SEDs relatively simple.

``Slim" or ``thick" accretion discs have also been considered, and there is evidence, from theoretical models, that such discs are more 
appropriate for BHs with
 \Ledd$>0.3$ or so \cite[e.g.][and references therein]{Abramowicz88,Ohsuga11,Netzer13}. There are other ideas
that combine thin discs at large radii with thick, very hot X-ray producing structures closer to the BH \citep[e.g.][]{Done12}.
The complexity of such models require more sophisticated calculations, e.g. 2D radiative transfer and treatment of advection close to
the event horizon. This makes the comparison to observed SEDs more uncertain.

Fitting thin AD models to observed AGN spectra is an active field of research  which was reviewed in several papers: see e.g. \citet{Koratkar99} for works before 1999 and \cite{Davis11} for more recent work.
As discussed in \citet{Koratkar99}, most early attempts to fit thin AD models to observed SEDs
reached the conclusion that the SEDs predicted by the theoretical models are considerably
bluer than those observed. \citet{Blaes01} fit a thin AD spectrum to observed spectra of a single AGN, 3C 273, but also find that the model is bluer in the optical and that it underpredicts the near-UV emission. Larger AGN samples have been modeled by \citet{Shang05}, \citet{Davis07}, and \citet{Jin12}. \citet{Shang05} combine spectra from multiple sources, for 17 AGN, to span a similar wavelength range as the current work, and they find that their data roughly agrees with the thin AD model.
However, \citet{Davis07}, who measure the rest-frame far-UV and near-UV spectral slopes for a very large number 
(several thousand) of SDSS AGN and \citet{Jin12}, who model a sample of $\sim$50 SDSS AGN, confirm the earlier results cited in \citet{Koratkar99} that there are discrepancies between the thin AD model and observations. Both \citet{Shang05} and \citet{Davis07} discuss the effects of reddening due to dust within the source or the host galaxy, and \citet{Davis07} claim that this instrinsic reddening may be the cause of much of the discrepancy between the model and observations.

In addition to directly comparing thin AD models to AGN spectra, other work have used more indirect means of testing the thin AD theory. \citet{Bonning07} use $M_{BH}$ measurements to determine the characteristic accretion disc temperature for a sample of SDSS quasars and find that the observed colors do not follow the trend predicted by thin AD models for bluer colors at higher disc temperatures. Furthermore, \citet{Bonning13} find that the observed line intensities in SDSS AGN spectra do not show the expected trend for higher ionization at higher temperatures. However, they mention that taking into account disc winds gives better agreement between the observations and models.

Another way to check the validity of the thin AD theory is to compare measurements of the size of AGN ADs, via microlensing, to the size predicted by the thin AD theory. In general, such studies have found that ADs are larger than predicted by thin AD theory \citep[][and references therein]{Jimenez14}.

Given the mixed results of previous work, the status of thin AD theory in explaining the origin of AGN SEDs is unclear. However, all of the above studies are limited by relatively narrow wavelength coverage, by the possible variability between different observations taken by different instruments, and/or by stellar light contamination at long wavelengths. With our unique AGN sample and wide, single-epoch wavelength coverage, we are in the best position to test current thin AD theory.

This paper addresses the observed SEDs in a unique AGN sample at $z \simeq 1.55$ and their comparison with theoretical AD models.
In Section 2, we describe the sample selection, the observations, and the data reduction method. Section 3 describes the AD model that we use 
and the analysis of the observed spectra. Section 4 presents the results of fitting the thin AD model to the observed spectra and describes the additional assumptions necessary in those cases where the model does not adequately fit the data. Finally, in Section 5, we summarize our main conclusions from this work.
Throughout this work, we assume a $\Lambda$CDM cosmological model with $\Omega_{\Lambda}=0.7$, $\Omega_{m}=0.3$, and
$H_{0}=70\, {\rm km\, s^{-1}} \, {\rm Mpc}^{-1}$.

\section{Sample Observations and Data Reduction}
\label{data}

\begin{table}
  \scriptsize
  \caption{Summary of Observations and Data Reduction}
  \begin{tabular}{cccl}
    \hline
    Name & Date(s) Observed & $A_V^{(a)}$ & Notes \\
    \hline
    J1152+0702  & 2012 Apr 17 & 0.03 & \\
    J0155--1023 & 2011 Oct 24 & 0.06 & \\
    J0303+0027  & 2011 Oct 21 & 0.26 & \\
    J1158--0322 & 2012 Apr 15 & 0.08 & Adjusted VIS arm slope \\
    J0043+0114  & 2011 Nov 26 & 0.07 & Used Nov 24 standard star \\
    J0842+0151  & 2011 Dec 18 & 0.18 & \\
    J0152--0839 & 2012 Aug 10 & 0.08 & \\
    J0941+0443  & 2012 Mar 19 & 0.13 & \\
    J0934+0005  & 2012 Mar 01 & 0.11 & \\
    J0019--1053 & 2011 Nov 23 & 0.10 & \\
    J0850+0022  & 2012 Jan 23 & 0.13 & \\
                & 2012 Feb 22 &  & \\
    J0404--0446 & 2011 Nov 23 & 0.37 & \\
    J1052+0236  & 2012 May 17 & 0.13 & Adjusted VIS arm slope \\
                & 2012 May 20 &  & \\
    J0223--0007 & 2011 Nov 25 & 0.13 & \\
    J0136--0015 & 2011 Dec 02 & 0.08 & \\
    J0341--0037 & 2011 Dec 17 & 0.27 & \\
    J0143--0056 & 2012 Aug 10 & 0.08 & \\
                & 2012 Aug 15 &  & \\
    J0927+0004  & 2012 Feb 23 & 0.10 & \\
    J0213--0036 & 2012 Sep 12 & 0.11 & \\
    J1050+0207  & 2012 Feb 26 & 0.14 & \\
                & 2012 May 18 &  & \\
    J0948+0137  & 2011 Dec 18 & 0.41 & \\
                & 2012 Mar 19 &  & \\
                & 2012 May 11 &  & \\
    J1013+0245  & 2012 May 16 & 0.11 & \\
                & 2012 May 20 &  & \\
    \hline
    J0209--0947 & 2011 Dec 02 & 0.07 & \\
    J0240--0758 & 2011 Dec 22 & 0.07 & Used Dec 19 standard star \\
                & 2012 Jan 21 &  & \\
                & 2012 Sep 12 &  & \\
    J0213--1003 & 2011 Nov 26 & 0.09 & Used Nov 24 standard star \\
                & 2011 Nov 29 &  & \\
    \hline
    J1108+0141  & 2012 May 17 & 0.13 & Adjusted VIS arm slope \\
    J1002+0331  & 2012 May 22 & 0.07 & Used May 20 standard star \\
    J0323--0029 & 2011 Oct 20 & 0.26 & \\
    J1005+0245  & 2011 Dec 19 & 0.10 & BALQSO \\
    J0148+0003  & 2012 Aug 19 & 0.11 & \\
    \hline
  \end{tabular}
  $(a)$ {Galactic extinction.}
  \label{tab:data}
\end{table}

\subsection{Sample Selection and Observations}
\label{sec:sample}

\begin{figure}
 \centering
 \includegraphics[width=80mm]{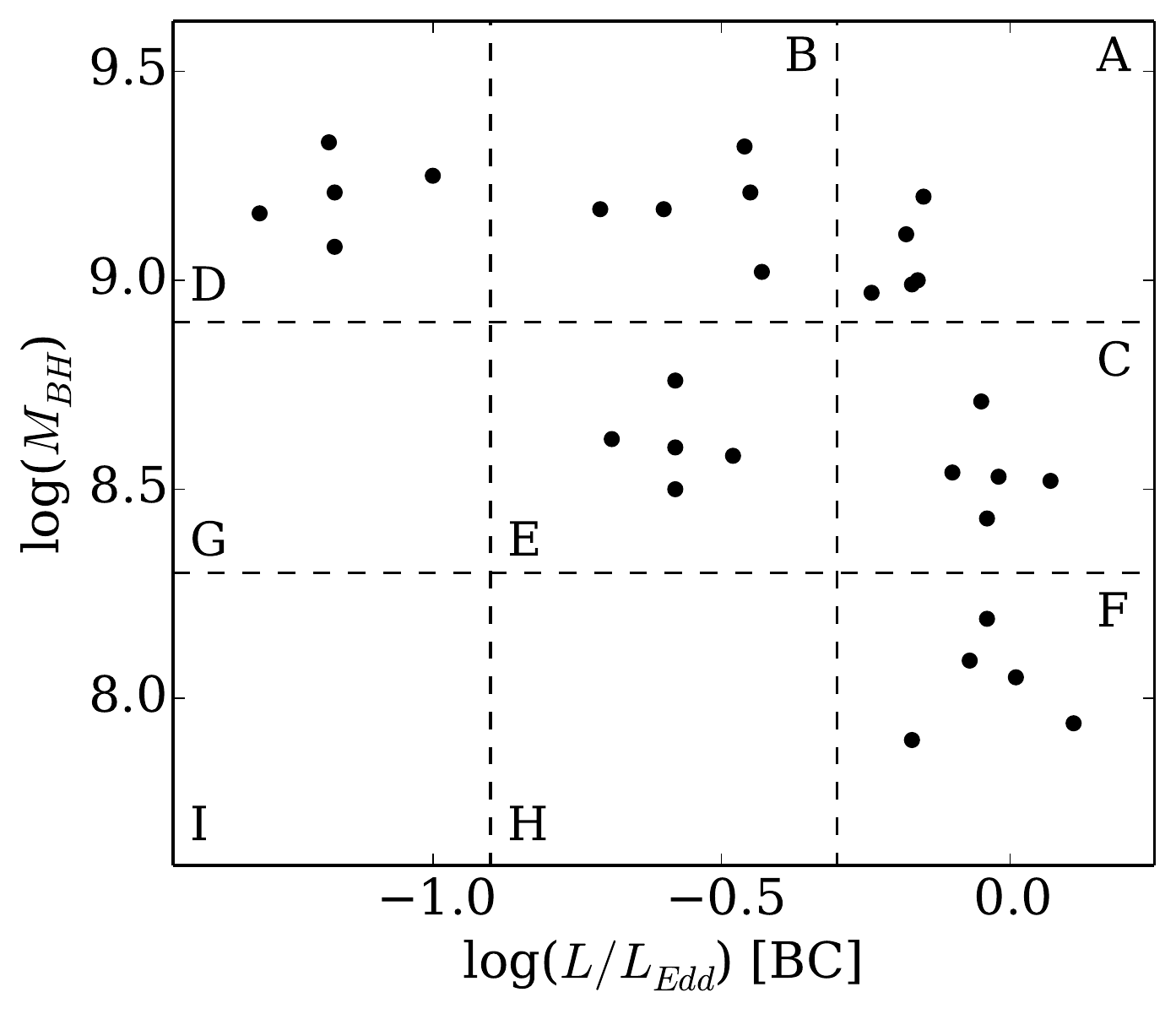}
 \caption{Our sample selection plotted on the $M_{BH}$--$L/L_{Edd}$ plane, using the measured values
   based on SDSS spectra and \citet{McLure04}.}
 \label{fig:mbh_ledd}
\end{figure}

We selected a sample of AGN from the seventh data release of the SDSS \citep{Abazajian09} that spans the widest possible range in $M_{BH}$ and $L/L_{Edd}$, within a narrow redshift range (z $\simeq$ 1.45 -- 1.65). We adopted this particular redshift range in order to include the most prominent emission lines $-$ \civ, \mgii, \Hbeta, and \Halpha\ $-$
all in one single-epoch spectrum, when observing with the {\it X-shooter} instrument at the VLT.
The sample was defined based on the values of $M_{BH}$ and $L/L_{Edd}$, estimated using measurements of the \mgii\ emission line in SDSS spectra, a standard bolometric correction factor, and relations given in \citet{McLure04}.
To evenly cover the $M_{BH}$$-$$L/L_{Edd}$ plane, we divide it into 9 bins, and select five objects per bin, as shown in Figure \ref{fig:mbh_ledd}.
We have currently observed the brightest 30 AGN from our sample, in bins A$-$F, 
with $M_{BH}$ ranging from $\sim$$2 \times 10^{8}$ to $4 \times 10^{9}$ M$_{\sun}$ and $L/L_{Edd}$ from $\sim$0.04 to 0.7. Observations are underway to add 9 more sources with $M_{BH}$ from $\sim 9 \times 10^{7}$ to $5 \times 10^{8}$ \Msun\ and $L/L_{Edd}$ from $\sim$ 0.05 to 0.3 (boxes G and H in Figure \ref{fig:mbh_ledd}).

The {\it X-shooter} instrument at the VLT splits 
incoming light into three arms, covering, simultaneously, the UV-blue (UVB), visible (VIS), and near-infrared (NIR) wavelength regions. 
Each arm consists of a prism-cross-dispersed spectrograph with its own optimized optics, dispersive element, and detector \citep{Vernet11}. 
Together, these three spectrographs produce a continuous spectrum from 3000 to 25000 \AA. At the redshift of our sample, this corresponds to 
a rest-frame wavelength range of $\sim$1200 to 9800 \AA. To minimize slit losses, we used the widest available slit widths 
of 1.6, 1.5 and 1.2'' in the UVB, VIS, and NIR arms, respectively. This provides a resolving power of about 3300, 5400, and 4300, respectively. The observations were taken under conditions where the seeing was $\leq$1'' and when the target was at an airmass of $\leq$1.4. 
To achieve accurate sky subtraction, the observations were split into sub-exposures and dithered along the slit. Table \ref{tab:data} lists the AGN sample and dates of observation.

We note that the fifth data release of the {\it Galaxy Evolution Explorer} ({\it GALEX}; \citealt{Morrissey07}) contains photometric measurements at effective rest-frame wavelengths of
$\sim$900 and 600 \AA\ for some of the AGN in this sample. While measurements at these wavelengths would be very useful for comparing to the thin AD model SEDs, the goal of this work is to make a careful comparison of observed single-epoch AGN spectra to thin AD spectra and these photometric measurements cover a wide bandpass (a width of over 1000 \AA\ at 2316 \AA\ in the observer frame).
Furthermore, the large elapsed time between the {\it GALEX} and {\it X-shooter} observations means that there is a large probability of spectral variability between the two datasets, especially at these short wavelengths.
Therefore, the use of {\it GALEX} data was not implemented.

\subsection{Data Reduction}
\label{sec:reduc}

The {\it X-shooter} spectra were reduced within the ESO Reflex environment \citep{Freudling13}, using version 2.2.0 of the ESO {\it X-shooter} pipeline in nodding mode \citep{Modigliani10}, in order to 
produce absolute flux-calibrated, one-dimensional spectra. To briefly summarize the reduction routine, the detector bias and dark current were subtracted, and then the 
spectra were rectified and wavelength-calibrated.
To obtain both a relative and an absolute flux-calibrated result, we used the observation of a spectroscopic standard star from the same night as the AGN observation, 
or in just a few cases, a nearby night.

After running the {\it X-shooter} pipeline, we corrected the spectra for telluric absorption at $\sim$6900, 7250, 7650, and 8200 \AA, using the observation of a telluric standard star at 
a similar airmass as the AGN that was taken right before or right after the AGN observation,
with the same instrument setup as was used for the AGN observation.
For the wavelength region $\sim$8950 to 9800, there are stellar absorption features 
in the telluric standard star spectrum, so we instead used a model telluric spectrum that is adjusted to the resolution of our observations. We do not correct for telluric absorption 
in the NIR arm, and we instead removed the portions of the spectrum that are heavily affected by this absorption.

For all of the objects in the sample, there is very good agreement between the red end of the UV-blue spectrum and the blue end of the visible spectrum, and in most cases, 
there was also good agreement between the visible spectrum and the NIR spectrum. However, there are three cases where there was a clear mismatch between the visible and NIR
spectra output by the pipeline (J1052+0236, J1108+0141, and J1158-0322). For these 3
objects, we adjusted the slope of the visible
spectrum by dividing the spectrum by the SDSS spectrum, fitting a low-order polynomial to
the result, and then dividing the {\it X-shooter} spectrum by this polynomial.

Finally, we corrected the spectra for Galactic extinction, using the maps of \citet{Schlegel98}
and the \citet{Cardelli89} extinction law. The values of $A_V$ due to the Galaxy range from 0.03 to 0.4 (see Table \ref{tab:data} for the $A_V$ value for each object).

\begin{figure*}
 \centering
 \includegraphics[width=160mm]{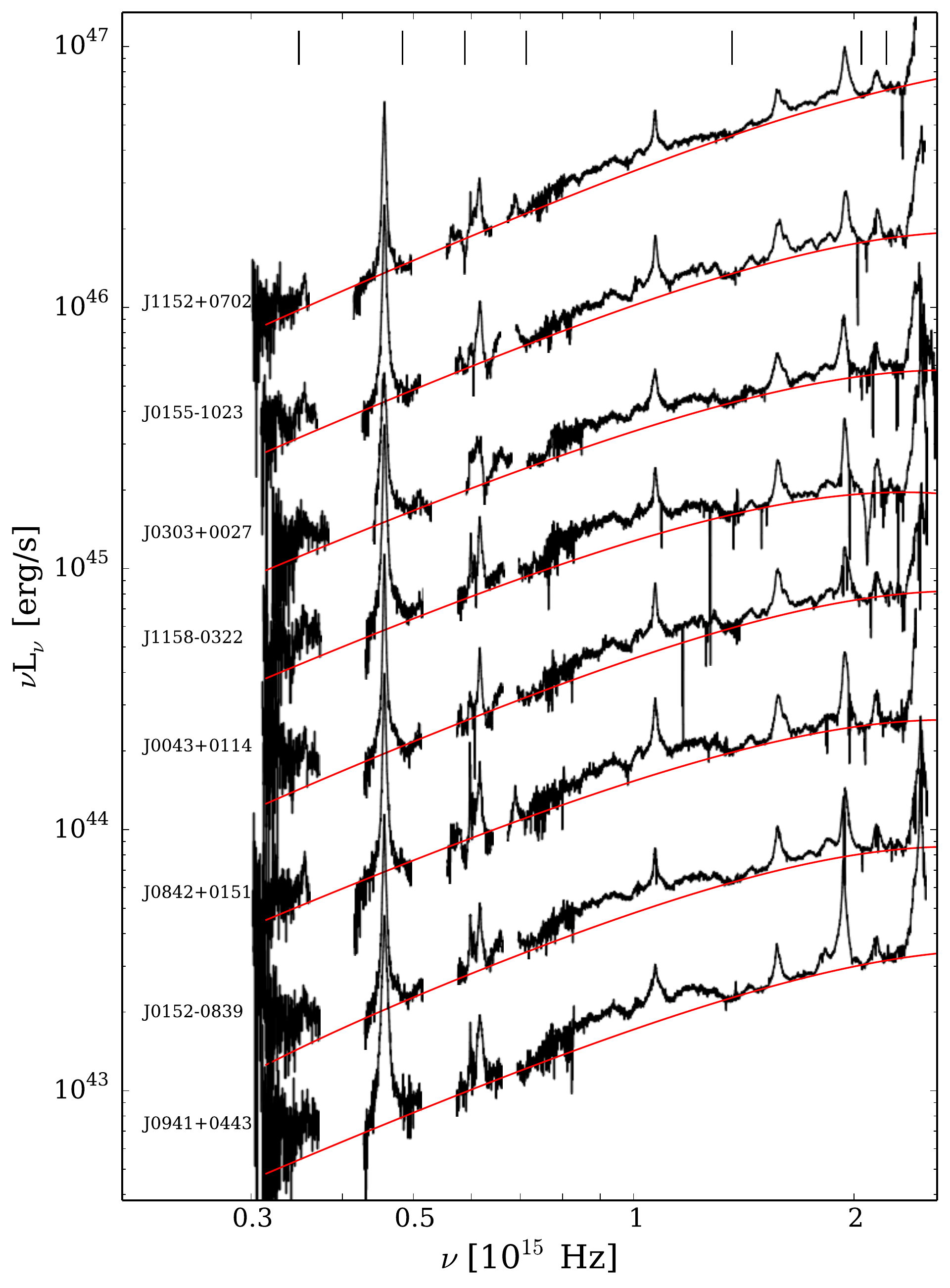}
 \caption{Spectra with the best-fit thin AD models (red curves) over-plotted. The spectra are
   multiplied by a constant for display purposes. This figure shows the 22 objects with satisfactory thin AD model fits, without correcting for intrinsic reddening or disc winds.
   The objects are ordered by source luminosity, as determined from $\lambda L_{\lambda}(3000)$\AA.
   The vertical lines at the top of each panel indicate the continuum regions used for fitting the models to the spectra.
   }
 \label{spectra1}
\end{figure*}
\addtocounter{figure}{-1}
\begin{figure*}
 \centering
 \includegraphics[width=160mm]{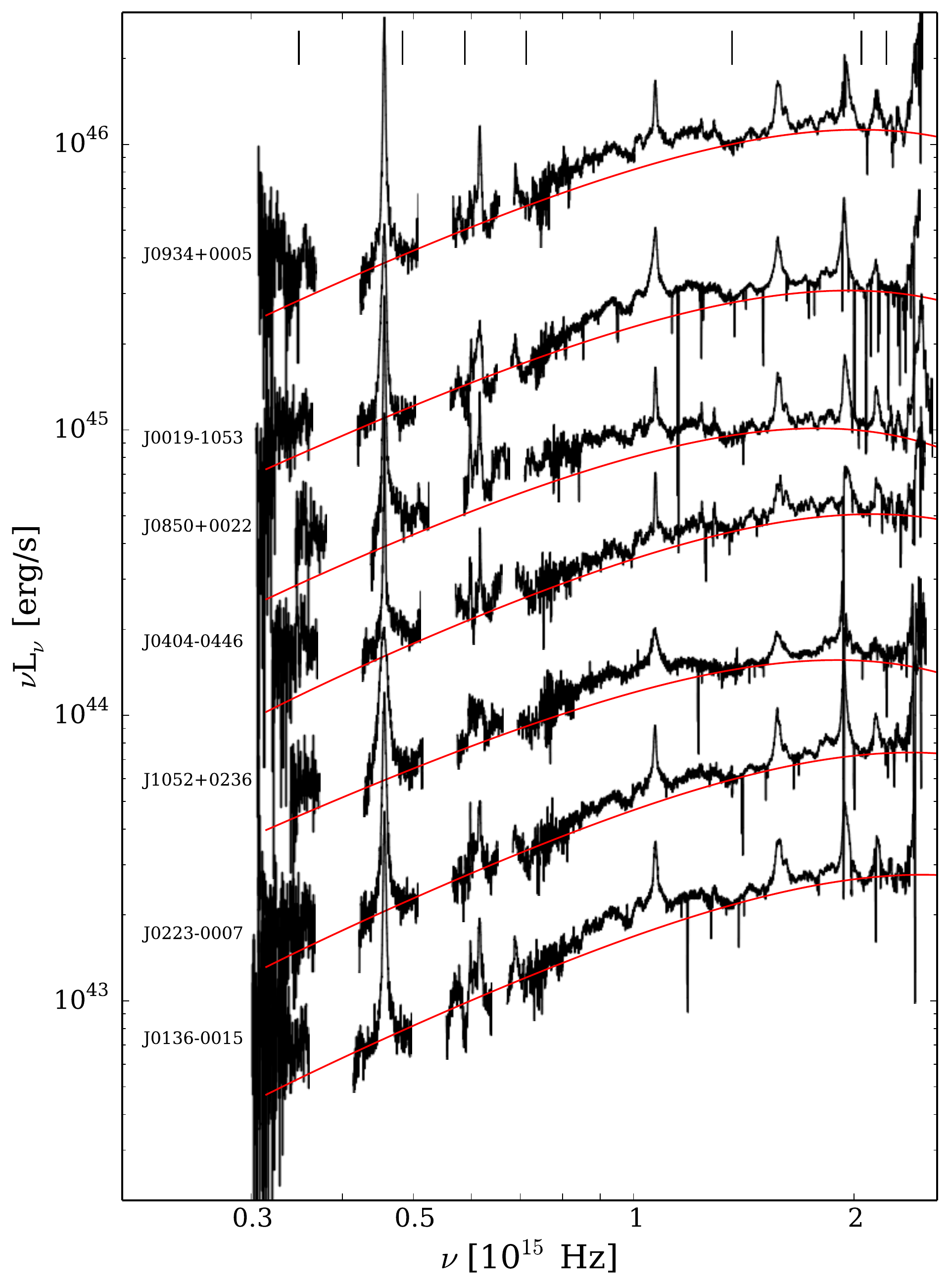}
 \caption{continued...}
\end{figure*}
\addtocounter{figure}{-1}
\begin{figure*}
 \centering
 \includegraphics[width=160mm]{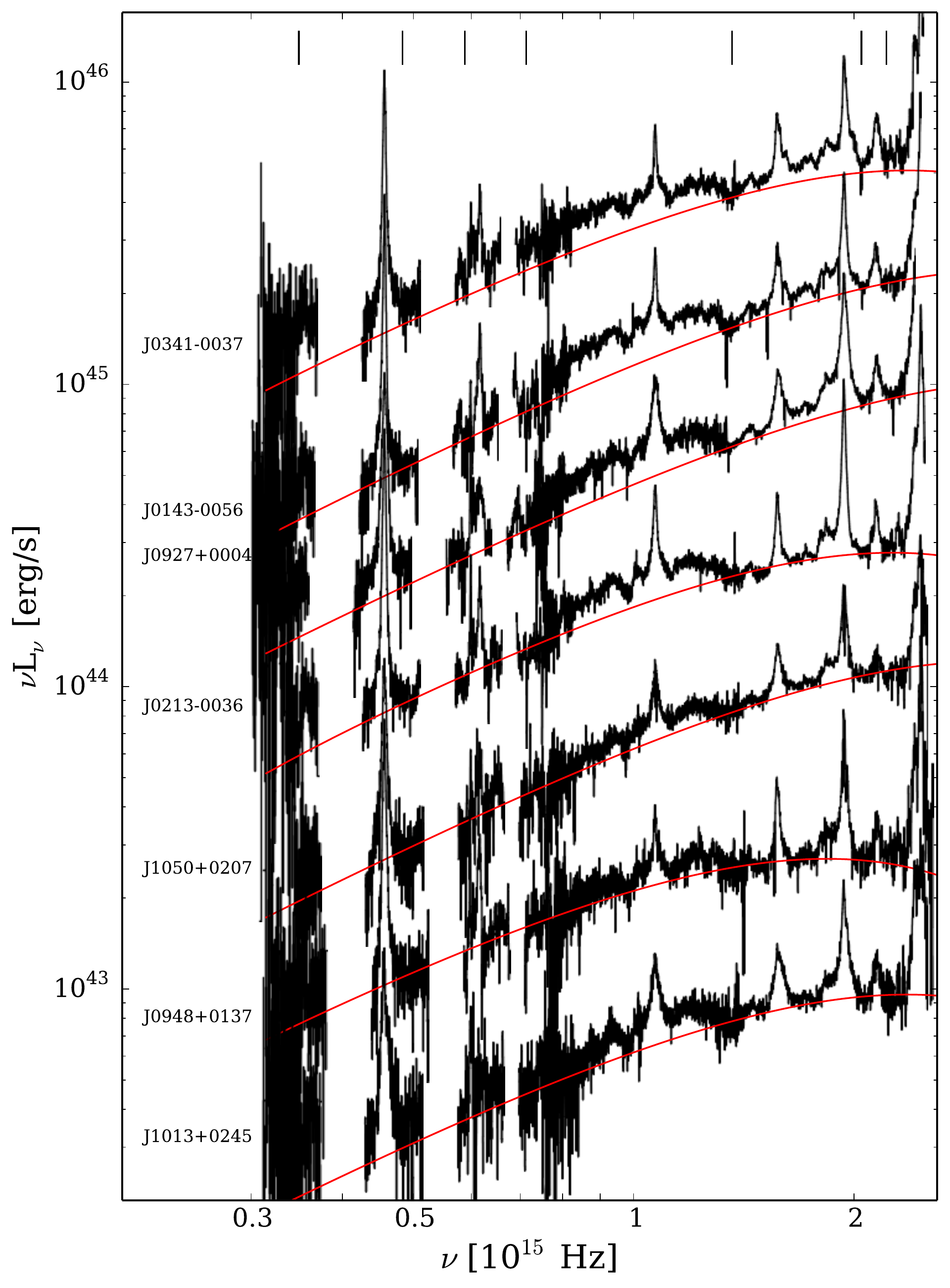}
 \caption{continued...}
\end{figure*}

\begin{figure*}
 \centering
 \includegraphics[width=160mm]{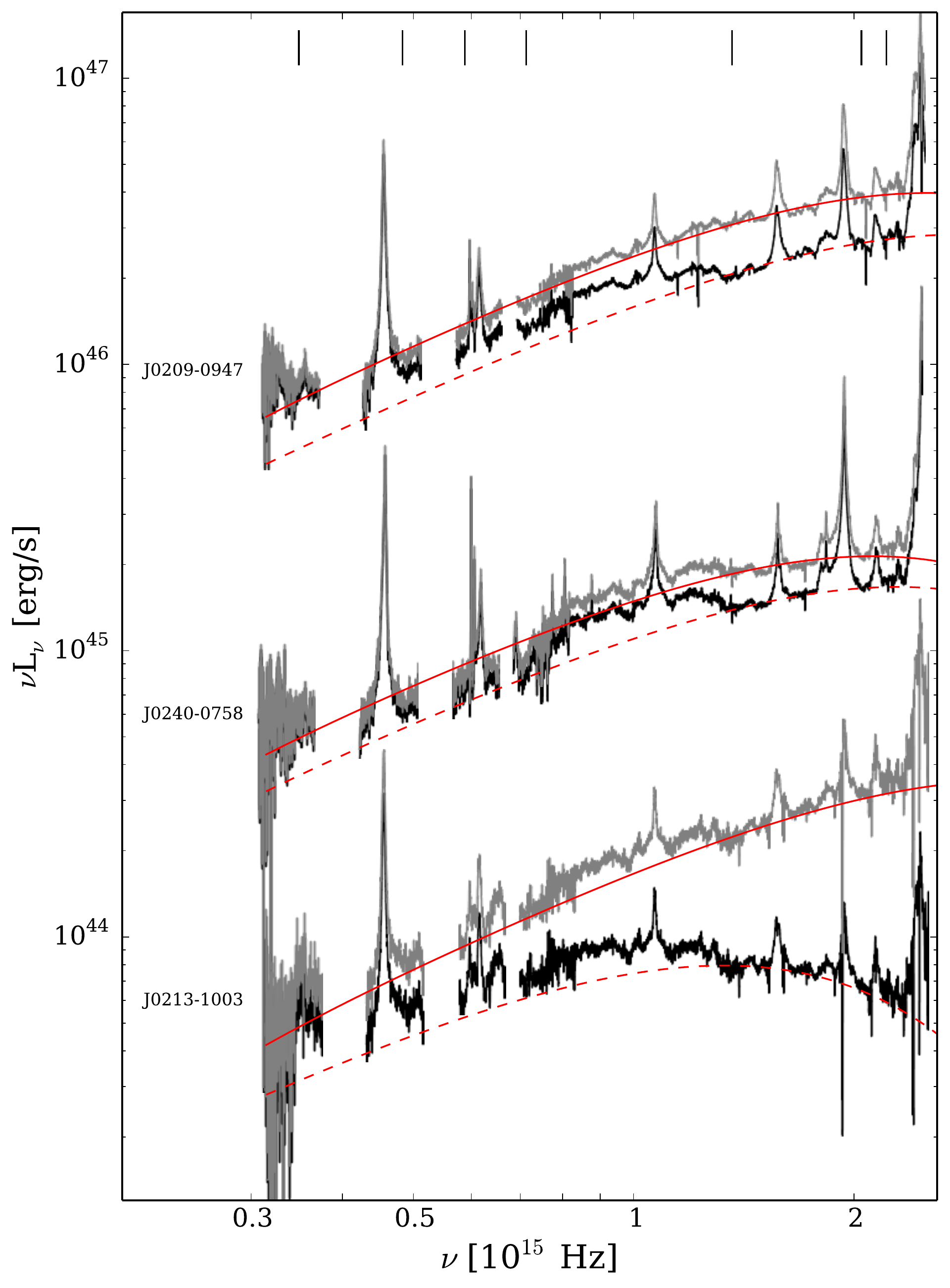}
 \caption{Same as Fig. \ref{spectra1}, but for those AGN which required an intrinsic
   reddening correction in order to obtain a satisfactory fit to the data. 
   The original spectra, with just a Galactic extinction correction, are shown in black, and the gray curves include the intrinsic reddening correction (see Table \ref{tab:master_table} for the extinction curves and values of $A_V$ used and Section \ref{res:red} for more details). The dashed red curve is the best fit to the original SED, and the solid red curve is the best fit to the dereddened SED.
   }
 \label{spectra2}
\end{figure*}

\begin{figure*}
 \centering
 \includegraphics[width=160mm]{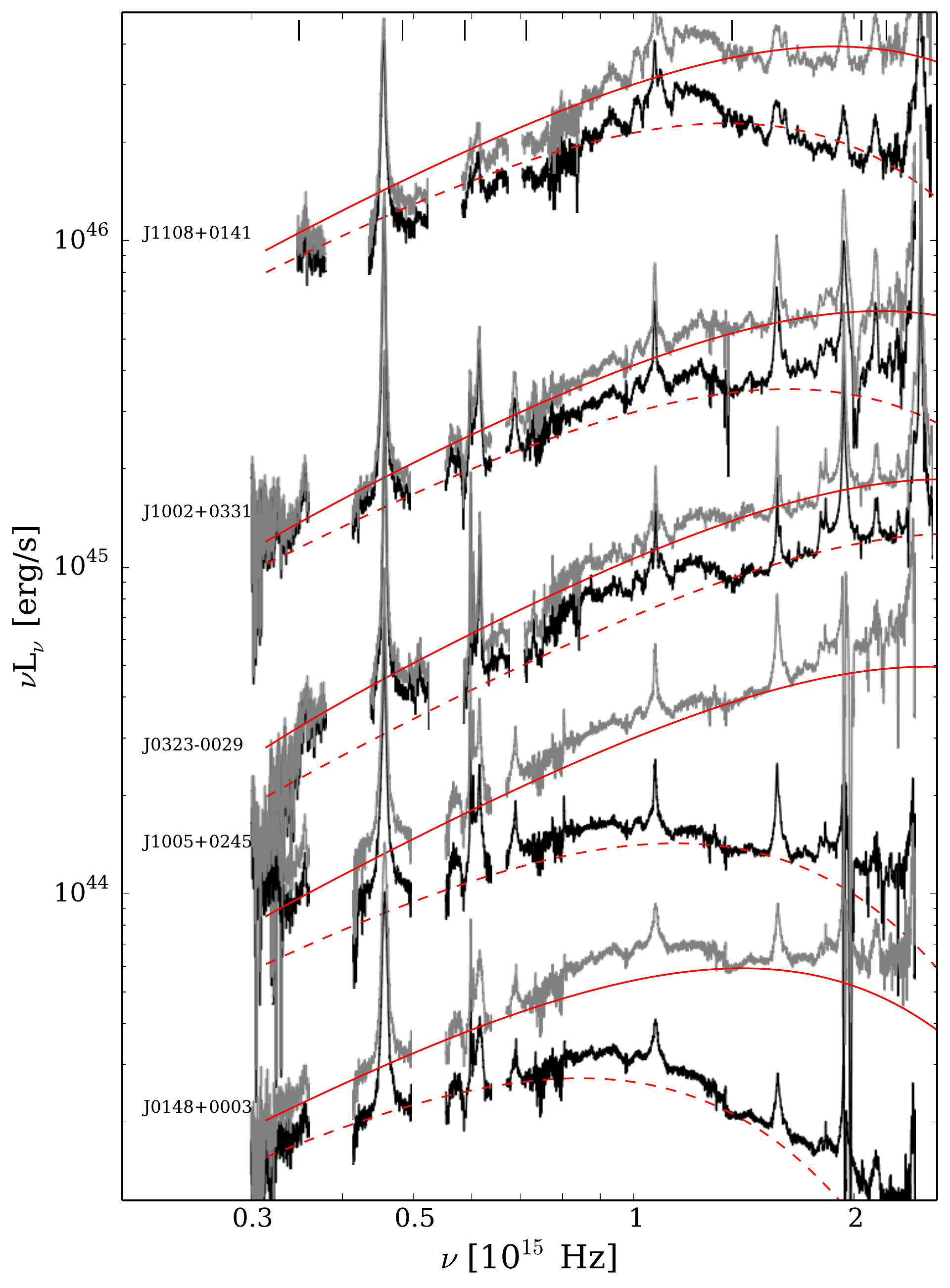}
 \caption{Same as Fig. \ref{spectra2}, 
   but for those AGN for which the best fit model to the dereddened spectrum provides only a marginal fit or does not fit the dereddened spectrum at all.
   }
 \label{spectra3}
\end{figure*}

Figs.~\ref{spectra1}, \ref{spectra2}, and \ref{spectra3} show the full SED spectra of all
30 sources. All spectra are corrected for Galactic extinction and some (Figs. \ref{spectra2} and \ref{spectra3}) are also corrected
for intrinsic reddening as described in Section \ref{res:red} below. The spectra are separated into these three figures based on the SED fitting results described in Section 4, and within each figure, they are ordered by source luminosity as determined from $\lambda L_{\lambda}(3000)$\AA. For consistency, the sources are ordered in this same way in Table \ref{tab:data}.

\section{Accretion Disc Models}
\label{models}

\subsection{Standard Thin Accretion Disc Models}
\label{sec:thin_AD}

As explained in Section 1, most current AD models are modified versions of the blackbody thin-disc model of SS73, with major improvements in two areas:
the inclusion of general relativity (GR) terms and the improvement of the radiative transfer in the disc atmosphere 
\citep[e.g.][]{Hubeny01,Davis11}.
In this work, we use the numerical code described in \cite{Slone12} to calculate the thin AD spectra. The calculations assume a SS73 disc with a variable viscosity 
parameter (chosen in this paper to be $\alpha=0.1$). As in all such models, the spin-dependent innermost stable circular orbit (ISCO) 
determines the mass-to-energy conversion efficiency, $\eta$, which ranges from 0.038 ($a_*$$=$$-$1) to 0.32 ($a_*$$=$0.998).
The calculations include Comptonization of the emitted radiation in the AD atmosphere and, for BH spin values of $a_*$ $\geq$ 0, full GR corrections. 
For retrograde discs with $a_*$ $<$ 0, the GR effects are not included, which is a fair approximation because of the large size of the ISCO 
($> 6 r_g$, where $r_g$ is the gravitational radius of the BH).

\subsection{$M_{BH}$ and $\dot{M}$ Determination}
\label{sec:mbh_mdot}

For the simplest thin AD calculations, the two input parameters are the BH mass ($M_{BH}$ in \msun) and the mass accretion rate (\Mdot\ in \msunyr). For each AGN in our sample, we calculate both $M_{BH}$ and $\dot{M}$ directly from the observed spectrum.

For $M_{BH}$, we use ``virial'' $M_{BH}$ estimates, which are fundamentally based on the results of reverberation mapping. From the observed spectra, we measure the FWHM of the \mgii\ emission line and the luminosity at 3000 \AA, and then we use the relations given in \cite{Trakhtenbrot12}. They find that the scatter in $M_{BH}$ estimates using the \mgii\ line, as compared to estimates using the H$\beta$ emission line, is $\sim$0.3 dex. We also estimate \Ledd\ using the luminosity at 3000 \AA\ and a luminosity-dependent bolometric correction (BC) factor, as described in \citet{Trakhtenbrot12}. These bolometric correction factors were designed to be consistent with the corrections of \citet{Marconi04}, and, for this sample, they range from 3.0 to 3.7. Throughout this work, we refer to these empirical estimates of \Ledd\ as \Ledd[BC].

To measure $\dot{M}$, we follow several earlier works based on the properties of thin ADs, e.g., \citet{Collin02,Davis11}. The SED of such systems,  at long enough wavelengths,
is given by a canonical power-law of the form $L_{\nu} \propto \nu^{1/3}$. Given a known \MBH, the mass accretion rate can be directly determined by using the 
monochromatic luminosity in the region of the continuum showing such a power-law. The only additional unknown is the disc inclination to the line of sight.
The expression we use here is taken from \cite{Netzer14} and is given by:
\begin{equation}
4 \pi D_L^2 F_{\nu} =f(\theta) [M_8 \dot{M_{\odot}} ]^{2/3} \left[ 
  \frac{\lambda}{5100{\rm \AA}} 
  \right]^{-1/3} 
   \,\, \ergcmsHz  \,\, ,
\label{eq:5100}
\end{equation}
where $F_{\nu}$ is the observed monochromatic flux, $M_8$ is the BH mass in units of $10^8\,\msun$,  \Mdot\ the accretion rate in units of \Msun/yr, and $D_L$ the luminosity distance.
The inclination dependent term, $f(\theta)$, gives the angular dependence of the emitted radiation. There are various possibilities for parameterizing this term
\cite[see][]{Netzer14}. Here we express it as: 
\begin{equation}
 f(\theta) = \frac{f_0 F_{\nu}}{F_{\nu}(\textrm{face-on})} = f_0 \frac {\cos \theta (1+ b(\nu) \cos \theta)}{1+b(\nu)}  \,\, ,
\label{eq:limb}
\end{equation}
where $\theta$ is the inclination to the line of sight and  $b(\nu)$ is a limb darkening function which, in the present work, we assume to be
frequency independent with  $b(\nu) = 2$. For this case,  $f_0 \simeq 1.2 \times 10^{30} ~ {\rm erg/sec/Hz}$. This constant was obtained from  realistic thin accretion models calculated by Slone and Netzer (2012), at a rest-frame wavelength of 5100 \AA. Because of this, it is slightly different from other values quoted in the literature (e.g. \citealt{Davis11}, as corrected in \citealt{Laor11}).

Throughout this work, we use \mdot\ to describe the normalized (or Eddington)  mass accretion rate, 
\begin{equation}
 \mdot= \frac{\Mdot}{\Mdot_{Edd}} \, ,
\label{eq:mdot}
\end{equation}
where $\Mdot = \Lbol/\eta c^2$ and $\Mdot_{\rm Edd} = L_{\rm Edd}/\eta c^2$.
We assume $L_{\rm Edd}=1.5 \times 10^{46}\,M_8\,\ergs$, which applies to solar composition gas. 
Using this definition, \mdot=\Ledd.

The above expressions can be re-arranged to estimate $\dot{M}$ by using the intrinsic $\lambda L_{\lambda}$ at a chosen wavelength. Longer wavelengths are likely to provide better estimates since the approximation is based on the $L_{\nu} \propto \nu^{1/3}$ part of the disc SED.
For \MBH\ $>10^8$\msun\ and relatively small \Mdot, this section of the SED corresponds to wavelengths longer than about 6000\AA\ \cite[e.g. figure 1 in][]{Netzer14}.
We choose the wavelength of $\lambda$$=$8600\AA, which is well beyond 6000 \AA, is located within the K-band of the observed SEDs, and is clear of emission lines.

A disadvantage associated with this choice is that towards longer wavelengths, the stellar light in the host galaxy starts to contribute significantly to the measured continuum. This effect is much more significant for fainter AGN than those studied here \citep{Stern12}. However, even for an AGN with log $L_{bol}$ of 45.5 (ergs/s), the host luminosity is equal to one-third of the AGN luminosity at 7000 \AA, and, in the rest-frame J-band ($\sim$11 000$-$14 000 \AA), the host is as bright as the AGN \citep{Stern12}. Relations given in \citet{Elvis12}, which give the host luminosity as a function of \Lbol, \Ledd, and $z$, indicate that for the least luminous AGN in our sample (J1013+0245; log \Lbol\ = 45.8 [ergs/s]), the AGN luminosity is 60\% of the total luminosity in the rest-frame J-band. For the most luminous AGN studied here (J1152+0702; log \Lbol\ = 46.9 [ergs/s]), the AGN luminosity is 97\% of the total luminosity in the rest-frame J-band. The wavelength we use for calculating \Mdot, 8600 \AA, is at shorter wavelengths than the J-band and will thus have even less host galaxy contamination. However, if the AGN luminosity is providing as little as 60\% of the total observed luminosity at 8600 \AA, the estimate of \Mdot\ will be overestimated by as much as a factor of 2.

\section{SED Fitting}
\label{results}

\begin{figure*}
 \centering
 \includegraphics[width=125mm]{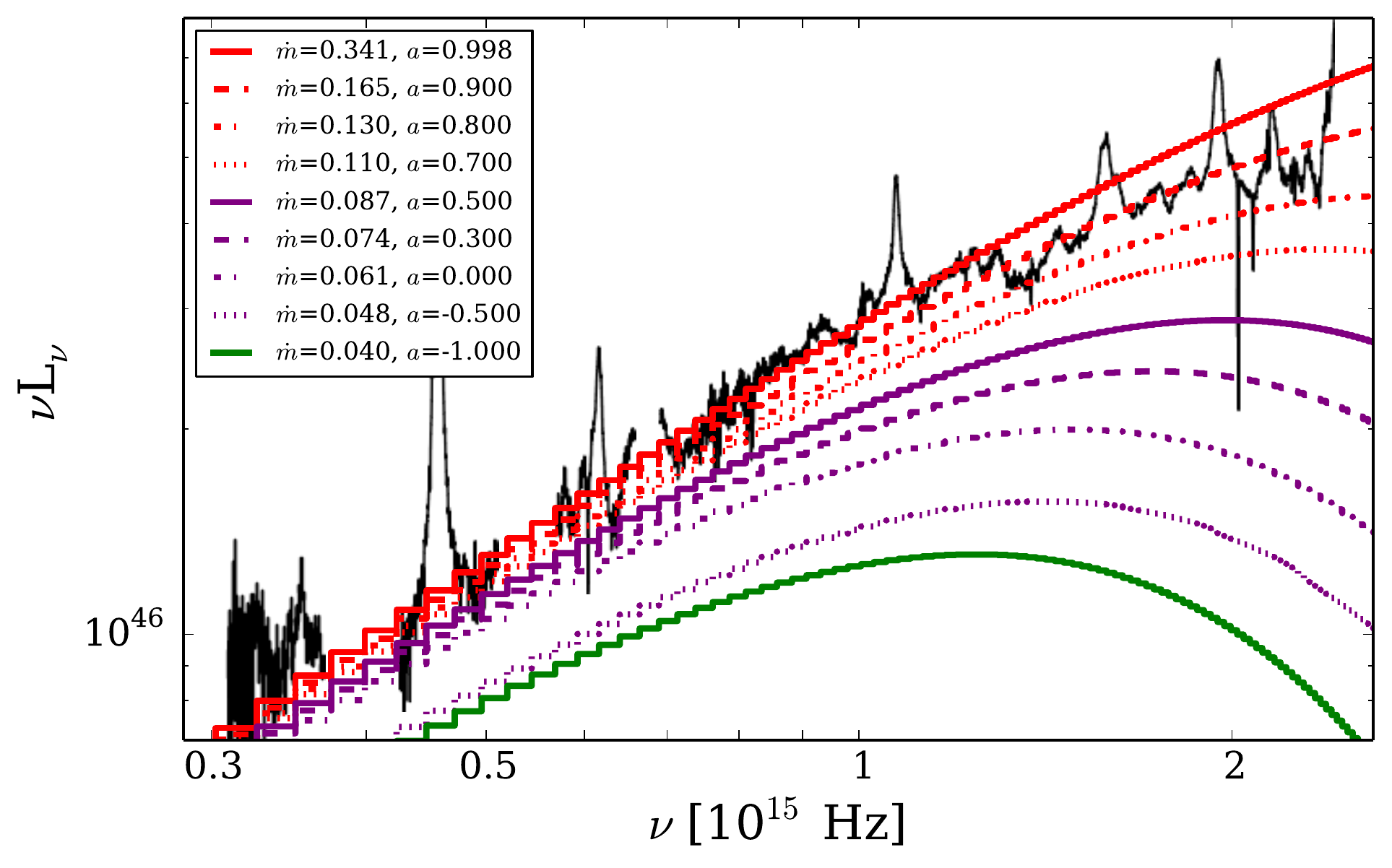}
 \caption{An illustration of the change in shape of a thin AD SED as a function of $a_*$
   (and therefore $\dot{m}$), with constant \mbh\ ($2.4 \times 10^{9}$ \Msun) and \Mdot\ (6.6 \Msunyr), for J0155-1023. With these measured values of $M_{BH}$ and $\dot{M}$, the best-fit model has \mdot=0.135 and $a_*$=0.820.
   }
 \label{fig:a_range}
\end{figure*}

\begin{figure*}
 \centering
 \includegraphics[width=125mm]{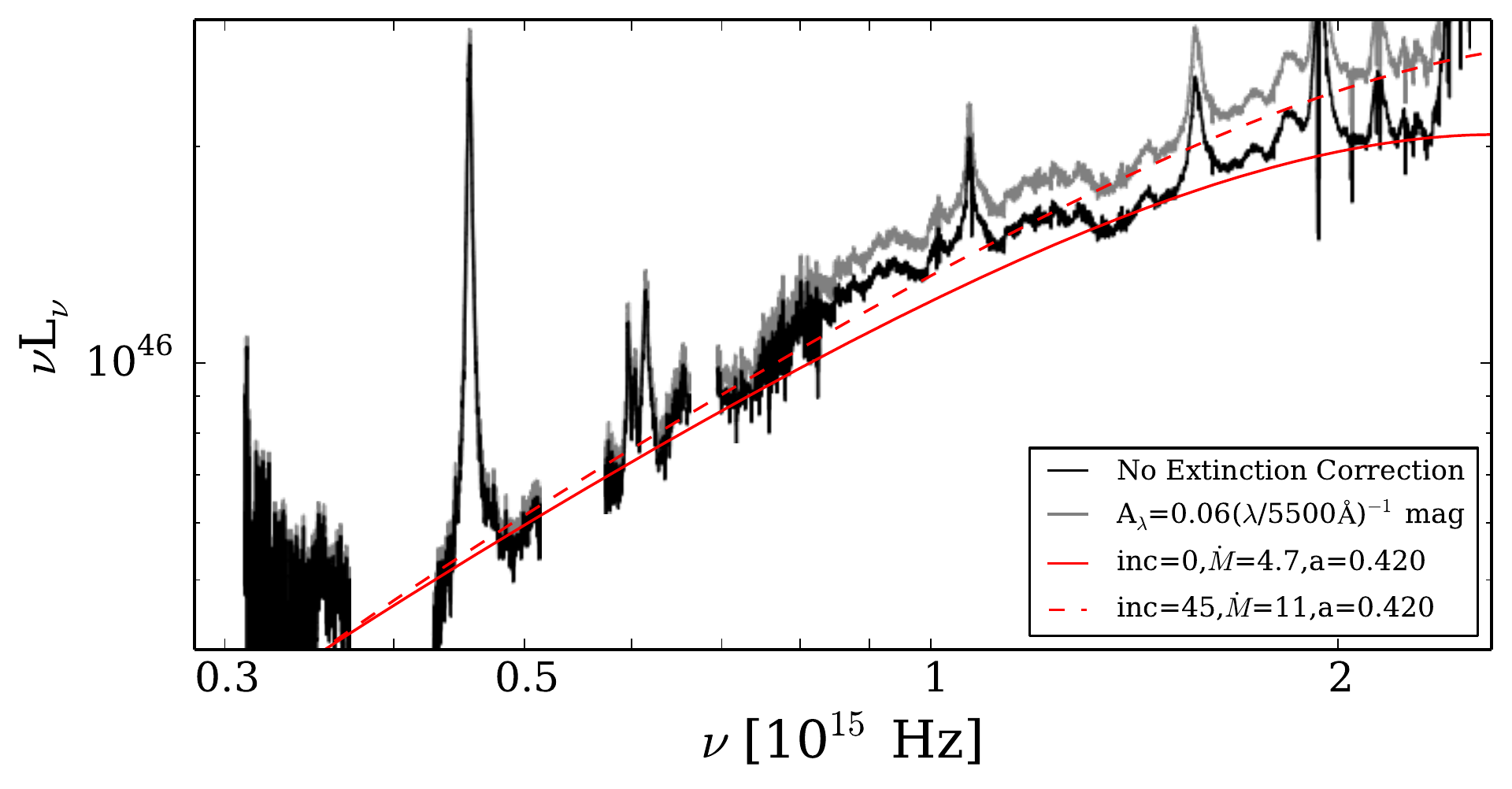}
 \caption{An illustration of SED variations due to the uncertainty in \Mdot\ caused by the
   unknown disk inclination. The solid red curve is for a face-on disc and the standard assumed \Mdot\ for J0152-0839. The dashed red curve is for a disc, with the same value of $M_{BH}$ and spin parameter ($a_*$), inclined 45$^{\circ}$ to the line-of-sight and hence having a larger \Mdot\ (in this case, a factor of 2.33 greater; see Eqn. \ref{eq:5100}). Including a correction for a small amount of instrinsic reddening (gray curve) results in a satisfactory fit between the model with a larger inclination (and \Mdot) and the data.
   }
 \label{fig:inc_red}
\end{figure*}

\subsection{Standard Thin AD SEDs}
\label{res:fitting}

The primary goal of this work is to determine what fraction of the AGN in our sample can be fit by the simple optically thick,
geometrically thin AD model. To test this, we use the \cite{Slone12} code described earlier to calculate a range of thin disc spectra.
For a preliminary analysis, we calculate a range of thin disc spectra for each source, using our measured values of $M_{BH}$, as determined from our \mgii-based measurement, and $\dot{M}$, as determined from the measured $\lambda L_{\lambda}$ at rest wavelength 8600\AA\ and the measured $M_{BH}$ (see Section \ref{sec:mbh_mdot}).
We assume a face-on disc and vary only the spin parameter.
Fig. \ref{fig:a_range} illustrates how the spin parameter, $a_*$, changes the shape of the thin AD SED, with \mbh\ and \Mdot\ held constant.

For this preliminary analysis, to find the best-fit model, and to evaluate the quality of the fit, we used a simple $\chi^2$ procedure, as has been done in previous works of this kind,
which is based on directly matching the data points to the models, in up to 7 line-free continuum windows.
These bands are centered on 1353, 1464, 2200, 4205, 5100, 6205, and 8600 \AA, with widths ranging from 10 to 50 \AA. 
For several objects at the higher end of the small redshift range of our sample, the 4205 and 5100 \AA\ windows are not usable because they 
fall within regions of strong atmospheric absorption. In these cases, the number of line-free windows is reduced.
Satisfactory fits are defined as those showing reduced $\chi^2 < 3$, and marginal fits are those with reduced $\chi^2 < 4.5$. For the error on each continuum point, we combine the standard error from Poisson noise and an assumed 5\% error on the flux calibration. We also allow for a simple scaling of the model.

For 20 out of the 30 AGN currently in our sample, the simple thin AD spectrum determined from the individually measured value of $M_{BH}$ and $\dot{M}$
provides a very good fit to the Galactic-extinction corrected observed SED.  For 2 additional sources (J0209-0947 and J0240-0758), the fit quality is marginal, and, for the remaining 8, 
there is no value of $a_*$ that produces even a marginal fit to the observed spectrum.

In this simple $\chi^2$-fitting procedure, however, we assume a face-on disc. This results in an uncertainty on \Mdot\ because our estimate depends on the inclination [\Mdot$\propto f(\theta)^{-3/2}$, see Eqn. \ref{eq:5100}]. For a given $M_{BH}$ and $a_*$, a larger $\dot{M}$ leads to a harder spectrum, which will affect our fitting of the SED at shorter wavelengths.
For example, Figure \ref{fig:inc_red} illustrates the effect of changing the inclination from face-on to 45$^{\circ}$, while keeping $M_{BH}$ and $a_*$ constant, for J0152-0839. In this case, the mass accretion rate for the 45$^{\circ}$ inclined disc is 2.33 times larger than for the face-on disc. As the figure shows, this can also be confused with intrinsic reddening of the source.

In order to take into account the errors in the input parameters, $M_{BH}$ and $\dot{M}$, as well as different inclinations of the observed disc, we carry out a more sophisticated statistical analysis using a Bayesian method.
For this, a grid of 103 950 models was constructed, again using the \citet{Slone12} code, for
evenly spaced values of $M_{BH}$, $\dot{M}$, $a_*$, and scaling factor, covering
7.70 $<$ log($M_{BH}$) $<$ 10.25, $-$1.50 $<$ log($\dot{M}$) $<$ +2.10, $-$1.0 $<$ $a_*$ $<$ +0.998,
and 1.000 $>$ $cos\theta(1+2cos\theta)/3$ $>$ 0.300 (see Table \ref{tab:bay}). Notice that
different values of the inclination $cos\theta$ only represent a different
scaling of a model as given by the term $cos\theta(1+2cos\theta)/3$ in Eqn. \ref{eq:limb}.

\begin{table}
\caption{Parameter values for the grid of AD models.}
\begin{tabular}{ccl}
  \hline
  Parameter & $\Delta$ & Min-Max values \\
  \hline
  $\log M_{BH}$                       &0.15      & $7.70:10.25$  \\
  $\log \dot{M}$                      &0.15      & $-1.50:+2.10$ \\
  $a_*$                                 &0.1       & $-1.0:+0.998$ \\
  cos$\theta$(1+2cos$\theta$)/3       &0.067     & $1.000:0.300$ \\
  \hline
\end{tabular}
\label{tab:bay}
\end{table}

For each model $m = m(M_{BH}, \dot{M}, a, cos\theta)$, we derived its
likelihood $(\mathcal{L}(m) \propto \exp(-\chi^2)/2)$, where $\chi^2 =
\sum (m_i - D_i)^2/\sigma_i^2$ is determined using the same 7 spectral windows
listed above, and $D_i$ is the average $\nu L_{\nu}$ in each window. There are no free parameters.

The posterior probability was then determined, for each of the 103 950 models,
as the product of the
likelihood $\mathcal{L}(m)$ and the priors on $M_{BH}$ and $\dot{M}$
(we have no prior knowledge on either $a_*$ or $cos\theta$).
We represent the priors as Gaussian distributions
centered on the observed values ($M^{obs}_{BH}$, $\dot{M}^{obs}$) and
with standard deviations ($\sigma_M$, $\sigma_{\dot{M}}$) given by their
uncertainties.
We have estimated $\sigma_{M}$ and $\sigma_{\dot{M}}$ to be 0.3 dex and 0.2 dex, respectively,
by a careful error propagation on the observed quantities. The errors associated to
$\sigma_{M}$ can be as large as 0.5 dex for a very conservative error
analysis that takes into account the uncertainties in the derivation
of $M^{obs}_{BH}$ from reverberation mapping results using the H$\beta$ emission line
and its scaling to the \mgii\ emission line. However, we found that
assuming $\sigma_M$ as high as 0.5 dex did not change the final
outcome of the analysis. In summary, our posterior probability is
given by:
\begin{eqnarray*}
  \textrm{posterior} \propto \exp(-\chi^2)/2) \times  \exp(-(M^{obs}\!-M^{mod})^{2}/2\sigma_{M}^{2}) \\
  \times \exp(-(\dot{M}^{obs}\! \times\! \frac{M^{obs}}{M^{mod}}-\dot{M}^{mod})^{2}/2\sigma_{\dot{M}}^{2}).
\label{eq:bay}
\end{eqnarray*}
Final probability distributions for each parameter were determined by
the marginalization (projection) of the posterior probability. See Appendix \ref{app:bayes}
for a full derivation of the posterior probability.

\begin{landscape}
\begin{table}
\caption{Measured and deduced physical parameters}
    \begin{tabular}{ccccccccccccccc}
    \hline
    Name & log(L3000) & log $M^{obs}_{BH}$ & log $\dot{M}^{obs}$ & log$L/L_{Edd}$ & log$M_{BH}$ & log$\dot{M}$ & cos$\theta$ & log $\dot{m}$   & $a_*$ & $\chi^2$/d.o.f.$^{(a)}$ & $<h \nu> ^{(b)}$ & EW(Ly$\alpha)^{(c)}$ & $A_{V}$ & Wind$^{(d)}$ \\
         & [erg/s] & [M$_{\sun}$] & [M$_{\sun}$/yr] & [BC] & [M$_{\sun}$] & [M$_{\sun}$/yr] &      &     &  &  & (Ryd) & (\AA) &      & \Mindot, \Moutdot\ \\
    \hline
    J1152+0702  & 46.549 &   9.24 &  0.98 & -0.38 &   9.51$^{+0.06}_{-0.06}$ &  1.19$^{+0.06}_{-0.06}$ & 0.56 & -0.23$^{+0.12}_{-0.31}$ &  0.998$^{+0.000}_{-0.032}$ & 1.6 &   2.45 & 120 &        &  \\
    J0155--1023 & 46.428 &   9.38 &  0.82 & -0.64 &   9.37$^{+0.39}_{-0.06}$ &  0.88$^{+0.07}_{-0.53}$ & 0.78 & -0.81$^{+0.37}_{-0.95}$ &  0.812$^{+0.158}_{-0.042}$ & 1.3 &   1.42 &  40 &        &  \\
    J0303+0027  & 46.372 &   9.57 &  0.56 & -0.88 &   9.80$^{+0.05}_{-0.05}$ &  0.15$^{+0.05}_{-0.05}$ & 1.0  & -1.56$^{+0.10}_{-0.29}$ &  0.998$^{+0.000}_{-0.032}$ & 2.9 &   1.40 &  37 &        &  \\
    J1158--0322 & 46.357 &   9.26 &  0.81 & -0.59 &   9.50$^{+0.06}_{-0.05}$ &  0.46$^{+0.05}_{-0.06}$ & 0.96 & -1.27$^{+0.15}_{-0.16}$ &  0.898$^{+0.035}_{-0.036}$ & 1.6 &   1.35 &  29 &        &  \\
    J0043+0114  & 46.272 &   9.28 &  0.72 & -0.69 &   9.21$^{+0.06}_{-0.06}$ &  0.75$^{+0.05}_{-0.05}$ & 0.83 & -0.86$^{+0.14}_{-0.13}$ &  0.707$^{+0.040}_{-0.039}$ & 2.3 &   1.42 &  39 &        &  \\
    J0842+0151  & 46.226 &   9.42 &  0.40 & -0.87 &   9.80$^{+0.05}_{-0.05}$ &  0.15$^{+0.05}_{-0.05}$ & 0.83 & -1.53$^{+0.08}_{-0.31}$ &  0.998$^{+0.000}_{-0.033}$ & 1.3 &   1.40 &  37 &        &  \\
    J0152--0839 & 46.116 &   9.01 &  0.67 & -0.56 &   8.90$^{+0.05}_{-0.05}$ &  1.05$^{+0.05}_{-0.05}$ & 0.73 & -0.63$^{+0.11}_{-0.11}$ & -0.600$^{+0.035}_{-0.034}$ & 0.8 &   1.41 &  37 &        &  \\
    J0941+0443  & 46.083 &   9.34 &  0.28 & -0.93 &   9.65$^{+0.05}_{-0.05}$ &  0.15$^{+0.05}_{-0.05}$ & 0.73 & -1.41$^{+0.10}_{-0.29}$ &  0.998$^{+0.000}_{-0.032}$ & 1.3 &   1.48 &  47 &        &  \\
    J0934+0005  & 45.939 &   8.91 &  0.59 & -0.64 &   9.13$^{+0.10}_{-0.13}$ &  0.70$^{+0.09}_{-0.17}$ & 0.56 & -1.01$^{+0.24}_{-0.42}$ &  0.273$^{+0.053}_{-0.631}$ & 1.5 &   1.30 &  23 &        &  \\
    J0019--1053 & 45.791 &   9.18 &  0.18 & -1.05 &   9.34$^{+0.06}_{-0.06}$ & -0.13$^{+0.31}_{-0.06}$ & 0.96 & -1.71$^{+0.43}_{-0.24}$ &  0.887$^{+0.043}_{-0.127}$ & 2.1 &   1.29 &  22 &        &  \\
    J0850+0022  & 45.742 &   8.57 &  0.83 & -0.48 &   8.90$^{+0.05}_{-0.05}$ &  0.45$^{+0.05}_{-0.06}$ & 0.87 & -1.27$^{+0.11}_{-0.12}$ & -0.897$^{+0.038}_{-0.038}$ & 2.6 &   1.26 &  18 &        &  \\
    J0404--0446 & 45.729 &   8.45 &  0.95 & -0.38 &   8.77$^{+0.06}_{-0.07}$ &  0.44$^{+0.06}_{-0.05}$ & 1.0  & -1.15$^{+0.23}_{-0.12}$ & -0.878$^{+0.571}_{-0.049}$ & 2.8 &   1.31 &  24 &        &  \\
    J1052+0236  & 45.722 &   9.46 & -0.24 & -1.39 &   9.74$^{+0.09}_{-0.76}$ & -0.39$^{+0.69}_{-0.09}$ & 0.78 & -2.24$^{+1.65}_{-0.67}$ &  0.960$^{+0.061}_{-0.746}$ & 1.0 &   1.28 &  20 &        &  \\
    J0223--0007 & 45.681 &   8.83 &  0.11 & -0.80 &   8.78$^{+0.11}_{-0.07}$ &  0.28$^{+0.08}_{-0.17}$ & 1.0  & -1.19$^{+0.40}_{-0.30}$ & -0.163$^{+0.782}_{-0.076}$ & 0.2 &   1.35 &  30 &        &  \\
    J0136--0015 & 45.650 &   8.92 &  0.08 & -0.92 &   8.79$^{+0.14}_{-0.08}$ &  0.31$^{+0.12}_{-0.08}$ & 0.92 & -1.15$^{+0.32}_{-0.24}$ & -0.065$^{+0.440}_{-0.073}$ & 0.8 &   1.37 &  32 &        &  \\
    J0341--0037 & 45.572 &   8.58 &  0.43 & -0.65 &   8.74$^{+0.07}_{-0.06}$ &  0.12$^{+0.22}_{-0.14}$ & 0.96 & -1.19$^{+0.32}_{-0.43}$ &  0.301$^{+0.120}_{-1.013}$ & 1.3 &   1.35 &  30 &        &  \\
    J0143--0056 & 45.534 &   8.62 &  0.18 & -0.73 &   8.69$^{+0.10}_{-0.21}$ &  0.03$^{+0.34}_{-0.21}$ & 0.96 & -1.07$^{+0.65}_{-0.68}$ &  0.679$^{+0.143}_{-1.279}$ & 0.6 &   1.49 &  48 &        &  \\
    J0927+0004  & 45.514 &   9.28 & -0.38 & -1.40 &   9.20$^{+0.05}_{-0.06}$ & -0.60$^{+0.06}_{-0.05}$ & 1.0  & -1.71$^{+0.12}_{-0.30}$ &  0.998$^{+0.000}_{-0.036}$ & 2.1 &   1.53 &  53 &        &  \\
    J0213--0036 & 45.487 &   8.77 &  0.00 & -0.93 &   8.72$^{+0.18}_{-0.14}$ &  0.24$^{+0.17}_{-0.25}$ & 0.87 & -1.18$^{+0.48}_{-0.57}$ & -0.173$^{+0.606}_{-0.817}$ & 0.3 &   1.33 &  27 &        &  \\
    J1050+0207  & 45.402 &   8.88 & -0.48 & -1.11 &   8.75$^{+0.19}_{-0.26}$ & -0.10$^{+0.40}_{-0.22}$ & 0.87 & -1.27$^{+0.88}_{-0.69}$ &  0.666$^{+0.248}_{-0.857}$ & 0.4 &   1.46 &  44 &        &  \\
    J0948+0137  & 45.317 &   8.34 & -0.15 & -0.64 &   8.73$^{+0.07}_{-0.08}$ & -0.15$^{+0.08}_{-0.07}$ & 0.92 & -1.51$^{+0.20}_{-0.17}$ &  0.096$^{+0.145}_{-0.116}$ & 0.6 &   1.27 &  19 &        &  \\
    J1013+0245  & 45.210 &   8.93 & -0.57 & -1.32 &   8.71$^{+0.19}_{-0.16}$ & -0.26$^{+0.28}_{-0.22}$ & 0.92 & -1.49$^{+0.61}_{-0.63}$ &  0.455$^{+0.318}_{-0.854}$ & 0.3 &   1.35 &  30 &        &  \\
    \hline
    J0209--0947 & 46.263 & 9.18 [9.25] & 0.89 [0.87] & -0.59 &   9.05$^{+0.05}_{-0.05}$ & 1.05$^{+0.06}_{-0.05}$ & 1.0  & -0.72$^{+0.12}_{-0.11}$ & -0.303$^{+0.037}_{-0.036}$ &  9.0 [1.8, 2.6] & 1.37 & 33 & 0.15$^{(e)}$  & 3.9, 7.8  \\
    J0240--0758 & 45.678 & 8.67 [8.72] & 0.49 [0.48] & -0.65 &   8.75$^{+0.06}_{-0.05}$ & 0.45$^{+0.05}_{-0.05}$ & 1.0  & -1.12$^{+0.11}_{-0.12}$ & -0.897$^{+0.035}_{-0.036}$ &  3.1 [1.9, 1.6] & 1.31 & 24 & 0.10$^{(e)}$  & 1.0, 3.1  \\
    J0213--1003 & 45.616 & 8.51 [8.71] & 0.90 [0.86] & -0.54 &   8.74$^{+0.07}_{-0.06}$ & 0.74$^{+0.08}_{-0.20}$ & 0.87 & -0.68$^{+0.34}_{-0.35}$ & -0.097$^{+0.654}_{-0.395}$ &  4.1 [2.4]      & 1.48 & 48 & 0.45  &   \\
    \hline
    J1108+0141  & 46.337 & 9.21 [9.31] & 0.90 [0.89] & -0.56 & & & & &   &  1.6 [2.9]       &  &  & 0.30  &  \\
    J1002+0331  & 46.195 & 8.93 [9.00] & 1.04 [1.00] & -0.41 & & & & &   &  4.5 [3.4, 2.5]  &  &  & 0.15$^{(e)}$  & 3.7, 11   \\
    J0323--0029 & 46.153 & 8.65 [8.72] & 1.11 [1.08] & -0.18 & & & & &   &  3.6 [3.6, 2.7]  &  &  & 0.15$^{(e)}$  & 4.3, 13   \\
    J1005+0245  & 46.062 & 8.86 [9.08] & 1.11 [1.08] & -0.47 & & & & &   &  11  [6.5]       &  &  & 0.45  &  \\
    J0148+0003  & 46.059 & 9.37 [9.58] & 0.59 [0.56] & -0.99 & & & & &   &  7.3 [3.6, 3.5]  &  &  & 0.45  & 2.7, 8.0  \\
    \hline
    \end{tabular}
    $(a)$ {Initial reduced $\chi^2$ before correcting for effects of intrinsic reddening or a disc wind. Within brackets is the $\chi^2$ parameter after correcting for instrinsic reddening and, in some cases, after correction for a disc wind.} \\
    $(b)$ {Mean energy of an ionizing photon in Rydberg.} \\
    $(c)$ {Ly$\alpha$ equivalent width, assuming a covering factor of 10\%.} \\
    $(d)$ {\Mindot\ and \Moutdot\ in M$_{\sun}$/yr.} \\
    $(e)$ {The Milky Way extinction curve provided for the best-fit model. For the other cases, a simple power-law with $A_V \propto \lambda^{-1}$ was used.}
  \label{tab:master_table}
\end{table}
\end{landscape}


This Bayesian analysis identifies which model has the highest probability of explaining the observed SED. However, to ensure that this model with the highest probability does indeed provide a satisfactory fit to the observed SED, we again use the same criteria above, based on the reduced $\chi^2$ statistic.
We find that the number of sources with a satisfactory fit (reduced $\chi^2$ $<$ 3) increases to 22. Previously, we could not find satisfactory fits for J0850+0022 and J0404$-$0446 in the simple $\chi^2$$-$fitting analysis, but by allowing more freedom in $M_{BH}$, $\dot{M}$, and inclination via the Bayesian method, we find that they can indeed be fit by the thin AD model.
Of the remaining 8 sources, 3 have a marginal fit (reduced $\chi^2$ $<$ 4.5; J0323$-$0029, J0240$-$0758, and J0213$-$1003), and 5 are not fit by the thin AD model.
Thus, we find that 73--83\% of the observed SEDs are consistent with an optically thick, geometrically thin AD model.
This fraction is very large considering earlier studies where most SEDs were not found to be consistent with AD spectra \citep[e.g.][]{Davis07,Jin12}.

The results for the sources with an SED that can be fit, satisfactorily, by the model with the highest posterior probability are shown in Fig.~\ref{spectra1}, and all model parameters, including updated values of $M_{BH}$, $\dot{M}$, and disc inclination, are listed in
Table~\ref{tab:master_table}. The values listed for $M_{BH}$, $\dot{M}$, $\dot{m}$, and $a_*$ are the median values given by the Bayesian procedure, and the errors are based on the range of parameter space that encloses 68\% of the probability distribution for each model parameter. The value of cos$\theta$ given corresponds to the best-fit model for each source. The dashed red curves in Figs. \ref{spectra2} and \ref{spectra3} are the models with the highest posterior probability for those sources without satisfactory fits to the observed SED.

We note that for many of the AGN where the thin AD model provides a good fit to the observed spectrum, there are small deviations from the local continuum
at some wavelengths. This is not surprising given the uncertainties on AD models, especially the radiative transfer in the disc atmosphere that
was not treated here in great detail \cite[see e.g.][]{Hubeny01}. However, the global fit is very good, and
the model adequately explains the overall shape of the SED for these sources.

We also find that the model spectrum that best fits the observed SED often under-predicts the observed luminosity in the observer-frame K-band, even for some of the 22 cases with overall satisfactory fits. One possibility for such a discrepancy is the contribution of stellar light from the host galaxy, as discussed in Section \ref{sec:mbh_mdot}.

There are several possibilities for the observed discrepancy between the thin AD model and the observed SEDs in the 8 cases where no satisfactory fit was found:
1) the AGN spectra could be affected by wavelength-dependent extinction
in the host galaxy (``intrinsic reddening'') that reduces, preferentially, the emitted radiation at shorter wavelengths.  
2) Disc winds can also preferentially suppress the shorter wavelength part of a thin AD SED \citep{Slone12}. 3) The AGN does not contain a thin AD.
We explore each of these possibilities below.

\subsection{Instrinsic Reddening}
\label{res:red}

\begin{figure*}
 \centering
 \includegraphics[width=140mm]{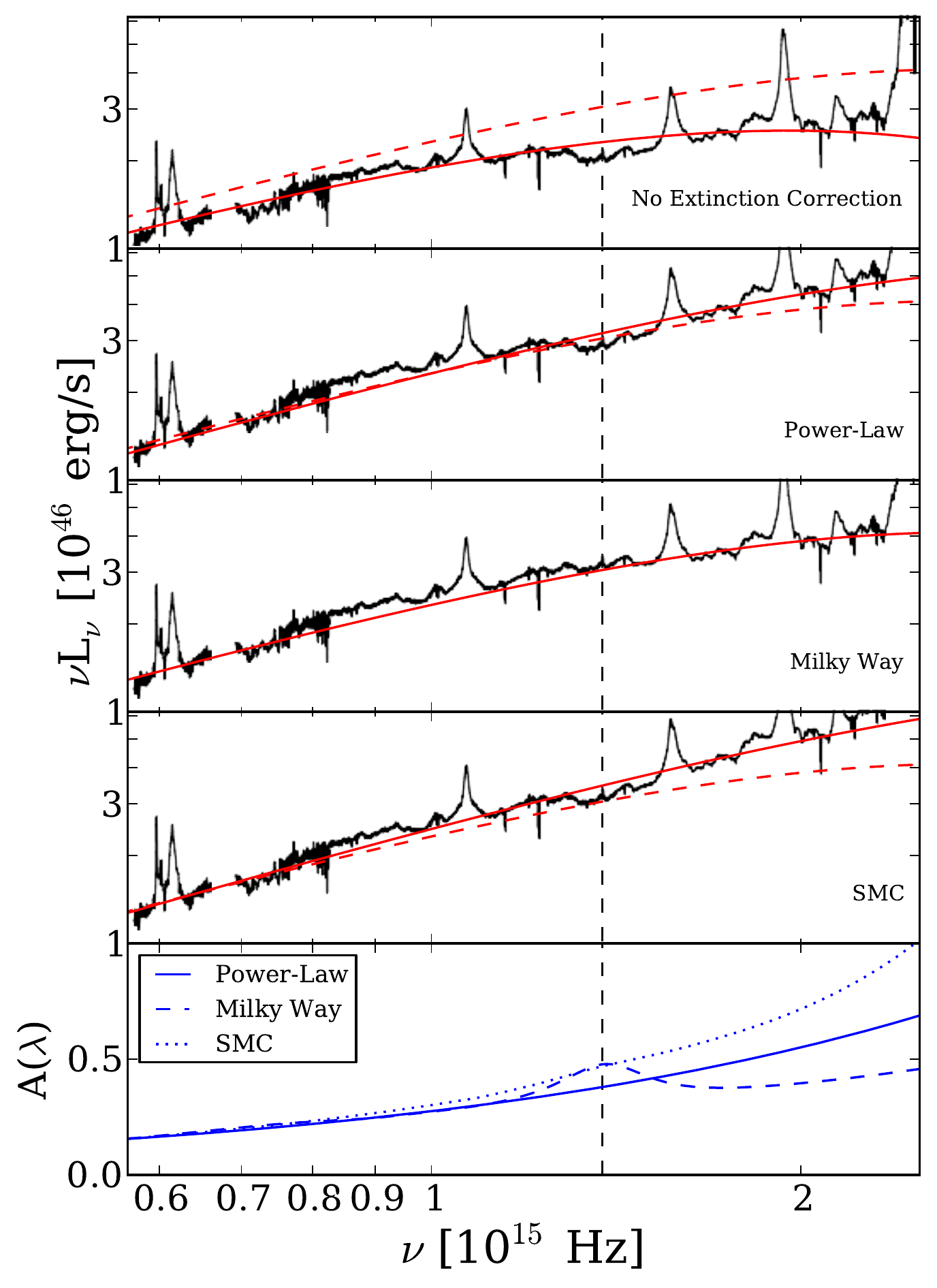}
 \caption{An example of one (J0209-0947) of the 4 AGN spectra that show evidence for the
   2175\AA\ bump in the Milky Way extinction curve (the vertical dashed line marks the location of 2175 \AA). The top panel shows the spectrum with no intrinsic extinction correction, and the next three panels show the spectrum corrected with a simple power-law, a Milky Way, and an SMC extinction curve, respectively, for $A_{V}$ = 0.15 mag. The solid red curve in each panel is the best-fit model for the uncorrected (top panel) and corrected (second to fourth panel) spectra, based on the simple $\chi^2$$-$fitting procedure. The dashed red curve is the best-fit model in the case of Milky Way extinction, overplotted for comparison in the other three panels. The three extinction curves used to correct the spectrum are plotted in the bottom panel.
}
 \label{fig:mw_red}
\end{figure*}

Reddening within the host galaxy (intrinsic reddening) can affect both the continuum shape and the broad emission lines in AGN spectra \citep{Netzer79,Netzer95}. \citet{Davis07} argue that such reddening may be the cause of
the discrepancy they find between their AD models and the spectral slopes they measure in SDSS spectra, whereas others \citep{Bonning07} do not find that taking into account reddening cures this discrepancy. 
We tested this possibility in our sample using various different extinction laws. We attempted to correct the observed spectra of the 8 AGN
for which the thin AD model either provides a marginal fit to the spectrum or does not fit the spectrum at all.
The extinction curves we try are (1) a simple power-law [$A(\lambda)=A_{o}\lambda^{-1}$], (2) the \citet{Cardelli89} Galactic extinction curve, and (3) the SMC extinction curve, as given in \citet{Gordon03}.

The most consistent approach would be to add this extinction as an additional parameter in the Bayesian analysis (i.e. modify Eqn. \ref{eq:bay}). However, our aim here is to illustrate simply whether an intrinsic reddening correction can cure the discrepancy between the observed and model SEDs for those 8 AGN that could not be fit in the previous section.
We defer the inclusion of these additional parameters in the Bayesian analysis to a later paper, when we have the full sample of 39 AGN.
We therefore use the initial, simple $\chi^2$-fitting, where we consider just the measured $M_{BH}$ and $\dot{M}$ values as input parameters to the models and vary only the spin parameter $a_*$, to determine which extinction curve and the amount of dereddening necessary for a satisfactory model fit.
For each extinction curve, we increase $A_{V}$, in increments of 0.05, until we find the value of $A_V$ that allows for the best fit to the observed SED.
We then update the value of $M_{BH}$ and $\dot{M}$ based on the dereddened spectrum, and we rerun the Bayesian analysis with these dereddened ``observed'' values of $M_{BH}$ and $\dot{M}$ ($M^{obs}_{BH}$, $\dot{M}^{obs}$) and the dereddened spectra for the 8 AGN.
The results are listed in Table \ref{tab:master_table}, including the reduced $\chi^2$ statistic for the best-fit model both before and after the intrinsic reddening correction, the extinction curve used, and the value of $A_{V}$. The median values of $M_{BH}$, $\dot{M}$, \mdot\ and $a_*$, listed in Table \ref{tab:master_table} for these sources, are based on the Bayesian analysis of the dereddened spectra. The dereddened spectra, with the best-fit models overplotted, are displayed in Figs. \ref{spectra2} and \ref{spectra3} as gray and solid red curves, respectively.

Out of the 8 AGN for which we applied an intrinsic extinction correction, 3 could be fit satisfactorily after the correction, and 4 marginally fit, with values of $A_{V}$ ranging from 0.10 to 0.45.
Therefore, after allowing for some moderate amount of instrinsic reddening, 25--29 out of the 30 AGN in our sample can be fit with a thin AD spectrum. This result shows convincingly that thin ADs are indeed the main power house of AGN.

Out of these 8 AGN for which we applied an instrinsic reddening correction, 4 were fit best by the thin AD model after applying the Milky Way extinction curve and assuming a moderate amount of extinction ($A_{V}$ = 0.10$-$0.15). As an example, in Fig. \ref{fig:mw_red}, we show the uncorrected spectrum for J0209-0947, and the spectrum after a simple power-law, a Milky Way, and an SMC extinction curve is applied. The spectrum with the Milky Way extinction curve applied clearly shows the best fit to the thin AD model. Without correcting for the strong 2175\AA\ bump that appears in the Milky Way extinction curve, it is not possible to obtain a satisfactory fit to the observed SED.

For the other AGN for which an intrinsic reddening correction cured the discrepancy between the observations and the models, a simple power-law extinction curve was sufficient. While there were several cases where using the SMC extinction curve allowed for an adequate fit to the model, there were no cases where the SMC curve allowed for a better fit than either the Milky Way or simple power-law curve.

It is possible that the other 22 AGN for which we found the thin disc model fit the observed SED without applying any intrinsic reddening correction are also affected by some intrinsic reddening. In particular, as mentioned in Section \ref{sec:mbh_mdot} and shown in Figure \ref{fig:inc_red}.
correcting the observed SED for a small amount of instrinsic reddening can have the same effect as increasing the value of $\dot{M}$ in the model. Therefore, there is some degeneracy between the effects of a small amount of intrinsic reddening and varying the value of $\dot{M}$.
In a later paper in this series, we will add the intrinsic reddening correction as another parameter in the Bayesian procedure in order to investigate this degeneracy.

\subsection{Disc Wind SEDs}
\label{res:winds}

\begin{figure*}
  \centering
  \includegraphics[width=135mm]{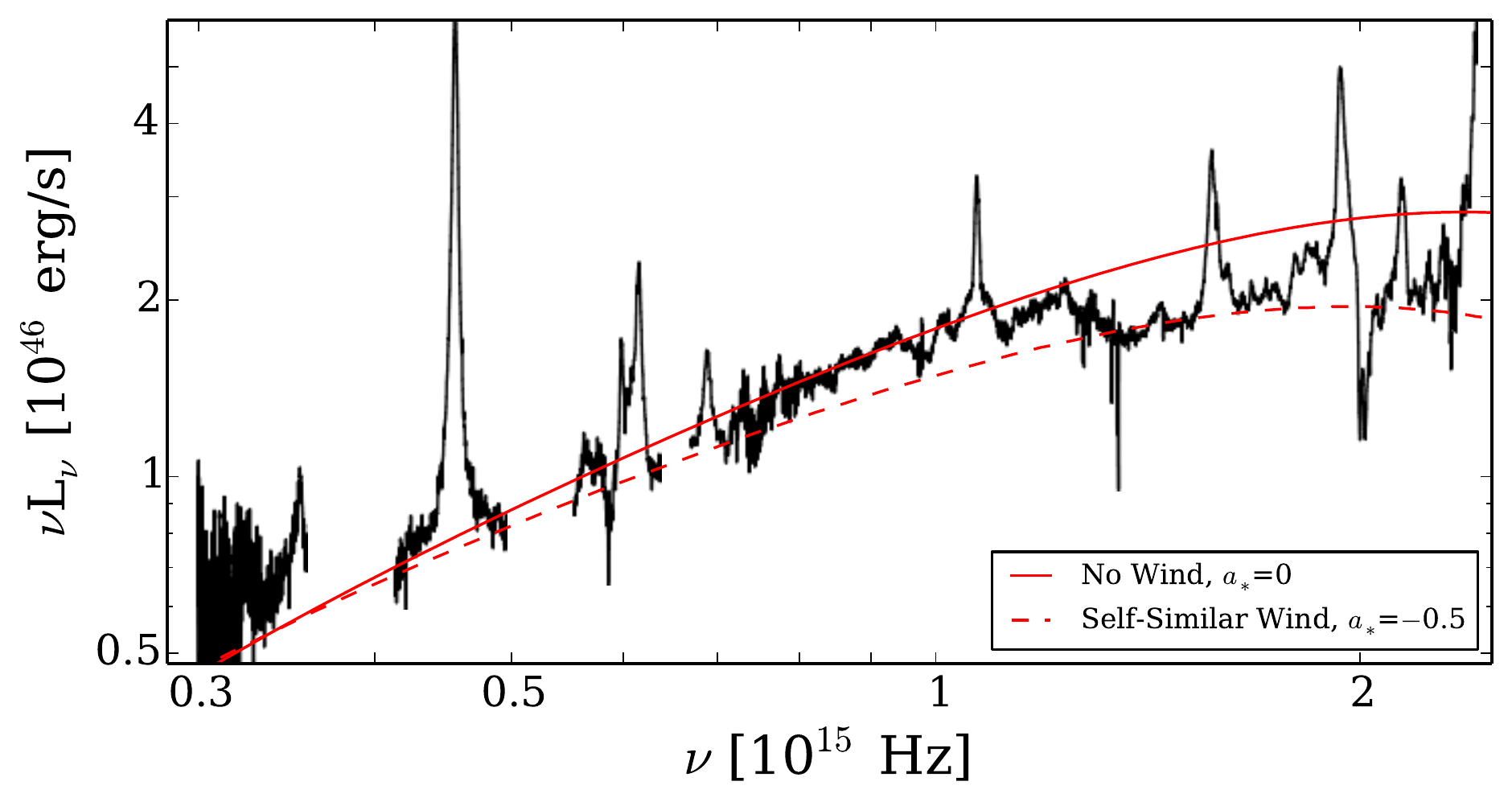}
  \caption{Example of a poor AD fit (solid red curve, corresponding to $a_*=0$), compared
    with a satisfactory fit when a disc wind is taken into account (dashed red curve, corresponding to $a_*=-0.5$ and $\dot{M}_{out}=3 \dot{M}_{in}$). In each case, the plotted curve corresponds to the spin value that gives the best fit to the observed SED.
  }
   \label{fig:disc_wind}
\end{figure*}

Numerous papers have discussed the possibility of mass outflows (``disc winds'') from the surface of the AD, and there is observational
evidence for the presence of such winds in high luminosity systems \citep[e.g.][and references therein]{Capellupo13,Tombesi13}.
The effect of disc winds on the observed SED was explored by \cite{Slone12}.
In this case, the mass accretion rate at the outer part of the disc ($\dot{M}_{out}$, at $r_{out}$), is 
larger than the mass accretion rate reaching the ISCO ($\dot{M}_{in}$ at $r_{in}$) by an amount which is determined by the wind properties and radial profile.
\cite{Slone12} presented three types of winds and explored the resulting modification of the SED compared with the case of no mass outflow
(the case of $ \dot{M}_{out} = \dot{M}_{in}$). In general, discs that have a mass outflow produce a softer SED for a given $\dot{M}_{out}$, compared with the
case of no disc wind. This is easy to understand because accretion in the innermost part of the disc contributes the most to the emitted short wavelength radiation
while accretion at larger radii is the main contributer to the long wavelength SED. \cite{Slone12} argued that this effect can explain the relatively flat (soft) 
SEDs of many AGN compared 
to the prediction of the thin AD model. A recent work by \citet{Laor14} discusses a different wind scenario where mass outflow from the inner part of the disc,
similar in nature to
stellar winds of massive stars, causes a similar change in the SED.

The \cite{Slone12} code provides the option of applying various disc-wind scenarios to the thin AD model. We have tested only one of the profiles discussed in \cite{Slone12}, the so called ``self-similar wind,'' where the mass outflow rate per decade of radii is constant.
To minimize the number of free parameters, we only consider three cases:
no disc wind (i.e. the standard thin disc discussed earlier), a case with $\dot{M}_{out}=2 \dot{M}_{in}$, and another with $\dot{M}_{out}=3 \dot{M}_{in}$.
As in Section \ref{res:red}, we apply this procedure only to those cases
where the simple thin AD model described in Section \ref{sec:thin_AD} does not provide a satisfactory fit to the observed spectra, and we keep $M_{BH}$ and $\dot{M}$ constant.
Obviously, some of these cases that are fit satisfactorily with a simple AD model can also be fit by a disc-wind model with somewhat different values of $\dot{M}$. However, as with instrinsic reddening, we do not add a disc wind as another parameter in the Bayesian analysis. We simply apply a disc wind in those cases that were not initially satisfactorily fit by the thin AD model to illustrate whether adding a disc wind can alleviate this discrepancy between the data and the model.

Of the 8 sources tested, we were able to find satisfactory fits to 4 AGN and a marginal fit to another 1 AGN, when including a disc-wind model. Note that we are fitting the disc-wind model to spectra that are not yet corrected for intrinsic reddening; we are testing here the disc-wind scenario as an alternative to the case of intrinsic reddening.
In all cases, we allow some freedom in the determination of $\dot{M}_{out}$, since the value determined from Eqn.~\ref{eq:5100} does not give the correct $\dot{M}_{out}$ in the presence
of a wind. Fig.~\ref{fig:disc_wind} shows one example where the presence of a wind improved 
considerably the agreement between model and observations.
In the 5 cases where adding a disc wind produced a model that at least marginally fit the observed SED, the $\chi^2$ values of the fit are similar to the $\chi^2$ values of the fit for the no-wind model to the dereddened SEDs (see Table \ref{tab:master_table}).
Thus, there is some degeneracy between correcting for intrinsic reddening and adding a disc wind to the model.

\subsection{Model Parameters and the ``Real'' $L/L_{Edd}$}

\begin{figure*}
 \centering
 \includegraphics[width=175mm]{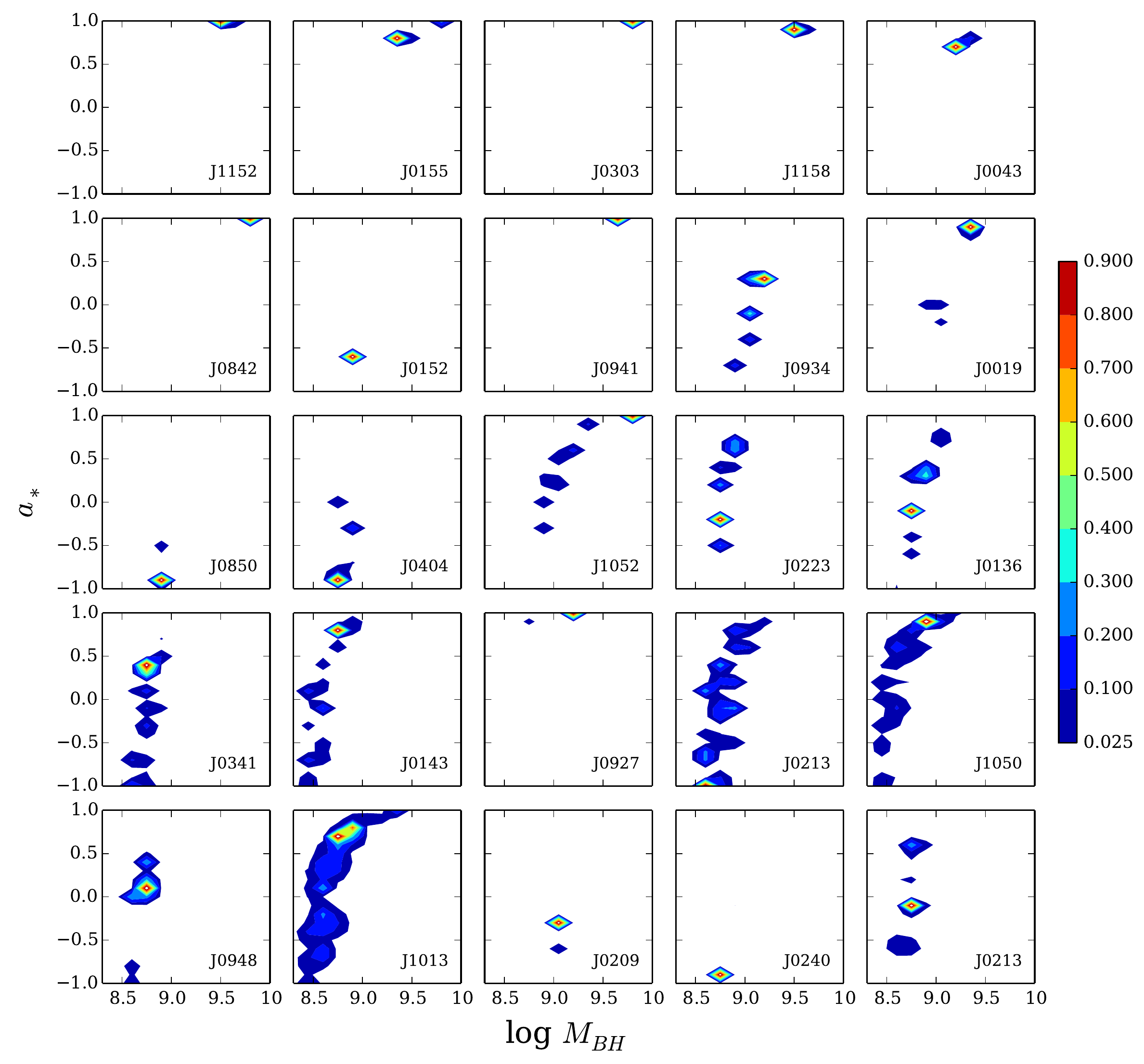}
 \caption{Contour plots of spin parameter $a_*$ versus $M_{BH}$ for sources with satisfactory fits
   without any instrinsic reddening correction (the first 22 panels) and for sources with satisfactory fits after correction for intrinsic reddening (the last 3 panels). The darkest blue contours correspond to a probability of less than 10 per cent.
 }
 \label{fig:a_mbh_grid}
\end{figure*}

\begin{figure*}
 \centering
 \includegraphics[width=175mm]{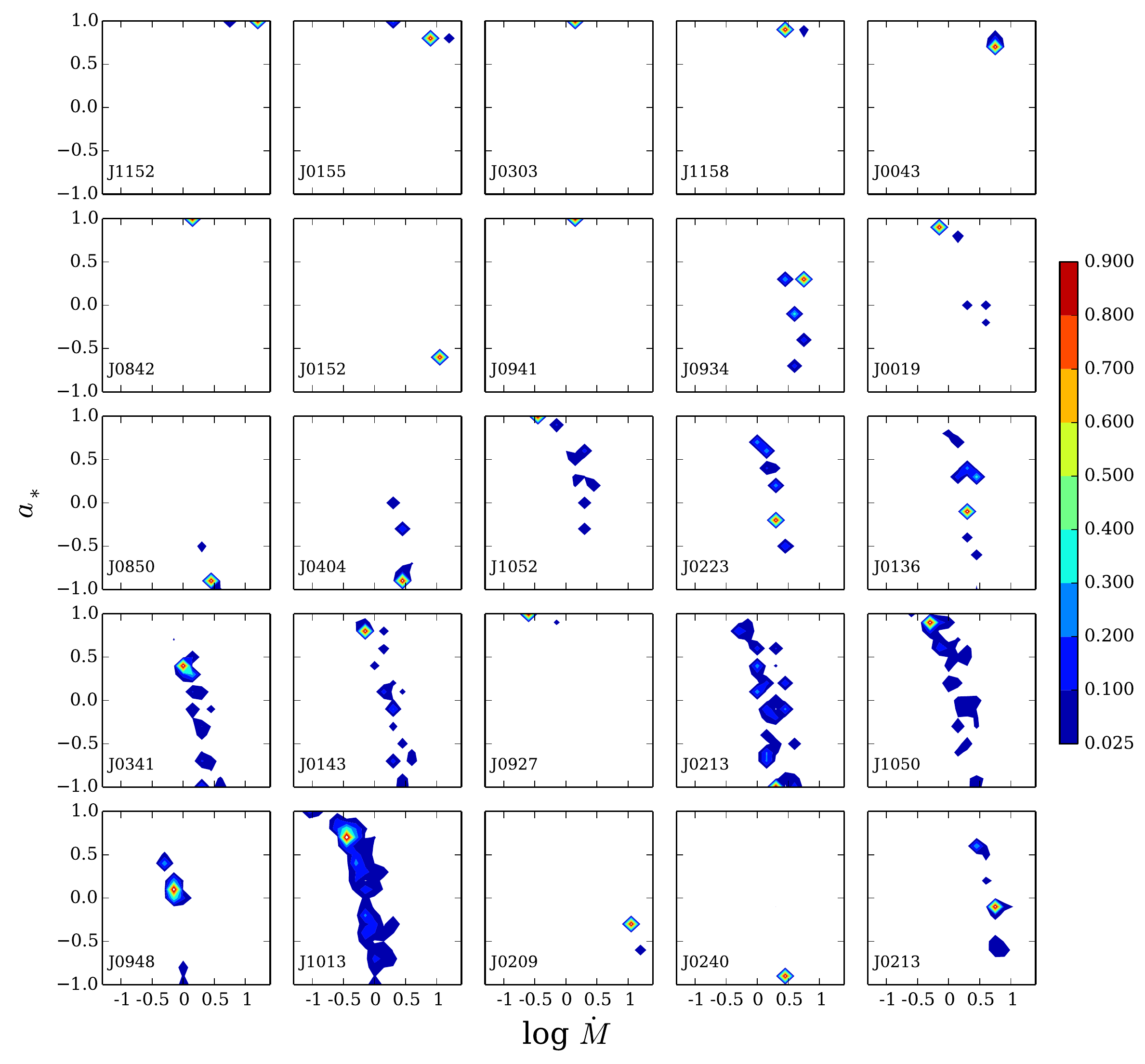}
 \caption{Same as Figure \ref{fig:a_mbh_grid}, but for $a_*$ and $\dot{M}$.}
 \label{fig:a_mdot_grid}
\end{figure*}

\begin{figure*}
 \centering
 \includegraphics[width=140mm]{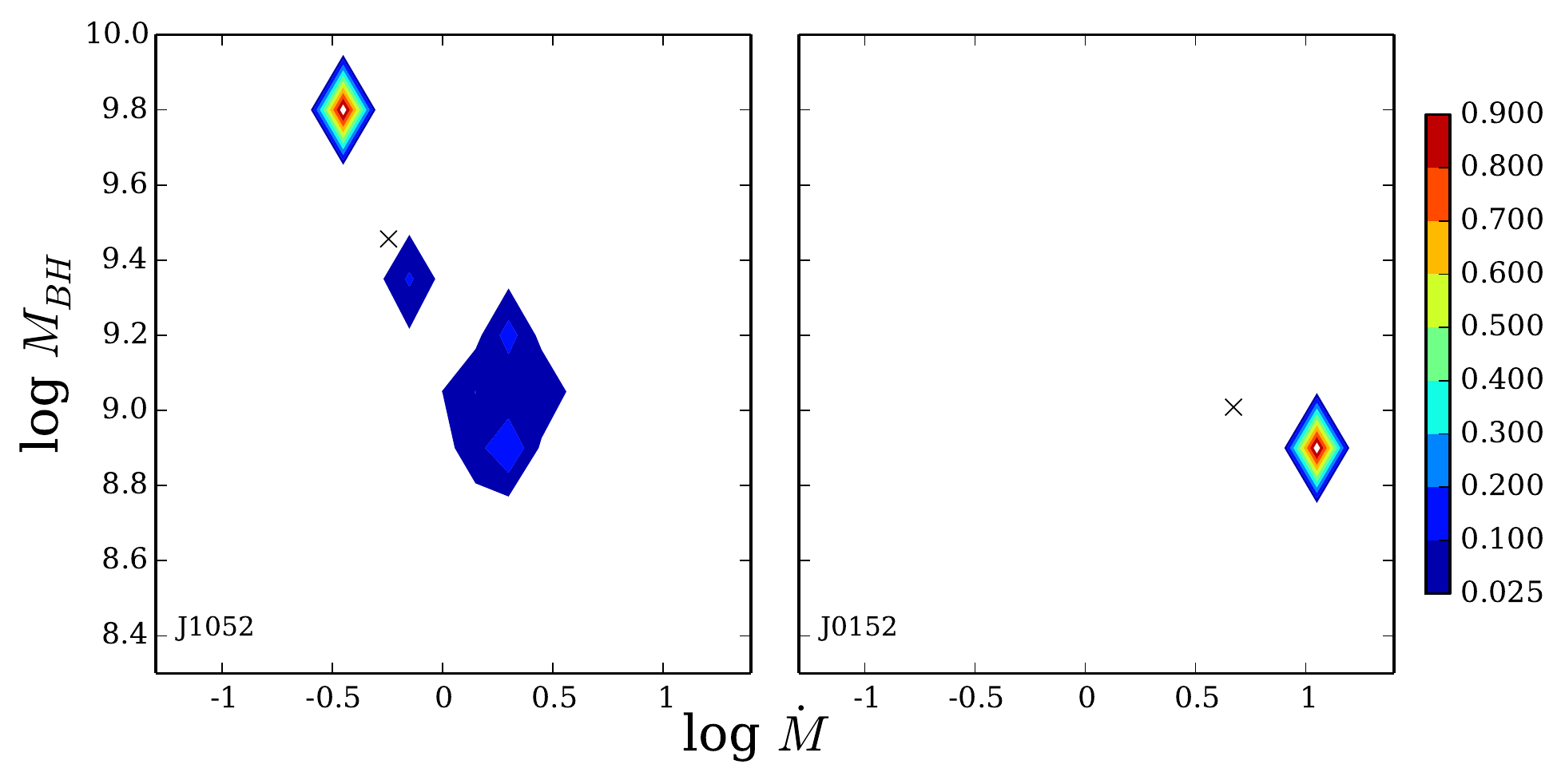}
 \caption{Contour plots of $M_{BH}$ versus $\dot{M}$ for two example AGN, one with a wide
   probability distribution for the two parameters (J1052+0236; left panel) and one with a narrow distribution (J0152$-$0839; right panel). For J1052+0236, the resulting probability distribution for the spin parameter $a_*$ is also wide, as shown in Figs. \ref{fig:a_mbh_grid} and \ref{fig:a_mdot_grid}, and for J0152$-$0839, it is narrow. The cross marks the observed values of $M_{BH}$ and $\dot{M}$, assuming face-on inclination.
 }
 \label{fig:mbh_mdot}
\end{figure*}

The primary input parameters for the thin AD models, besides the inclination, are $a_*$, $M_{BH}$, and $\dot{M}$. Figs. \ref{fig:a_mbh_grid} and \ref{fig:a_mdot_grid} show the probability contours for the spin parameter $a_*$ versus $M_{BH}$ and $\dot{M}$, respectively, for all AGN with a satisfactory fit. The last 3 panels in Figs. \ref{fig:a_mbh_grid} and \ref{fig:a_mdot_grid} are the 3 sources which required an intrinsic reddening correction for a satisfactory fit.
Fig \ref{fig:mbh_mdot} shows two example $M_{BH}$ versus $\dot{M}$ probability contours, one for an AGN with wide probability distributions for each parameter (J1052+0236) and one with narrow distributions (J0152$-$0839).

\begin{figure}
 \centering
 \includegraphics[width=80mm]{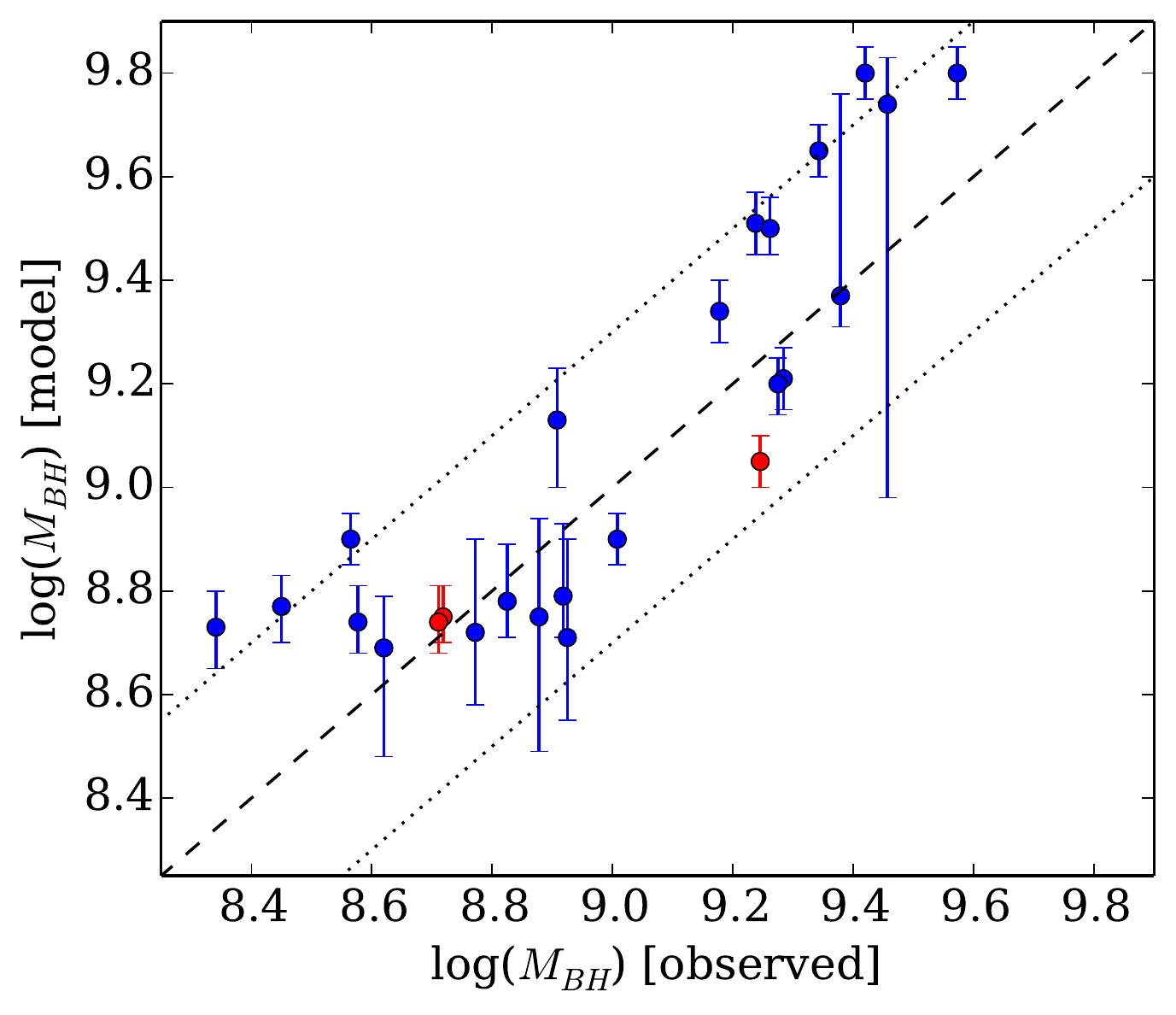}
 \caption{A comparison between the observed \mbh, measured directly from the spectra (Section
   \ref{sec:mbh_mdot}), and the median value of $M_{BH}$ from the Bayesian analysis. The red points are sources for which we applied an intrinsic reddening correction. For reference, the dashed line is the one-to-one line, and the dotted lines are $\pm$0.3 dex. The typical errors on log($M^{obs}_{BH}$) are 0.3 to 0.5 dex (Section \ref{res:fitting}).
   }
 \label{fig:mbh_comp}
\end{figure}

\begin{figure}
 \centering
 \includegraphics[width=80mm]{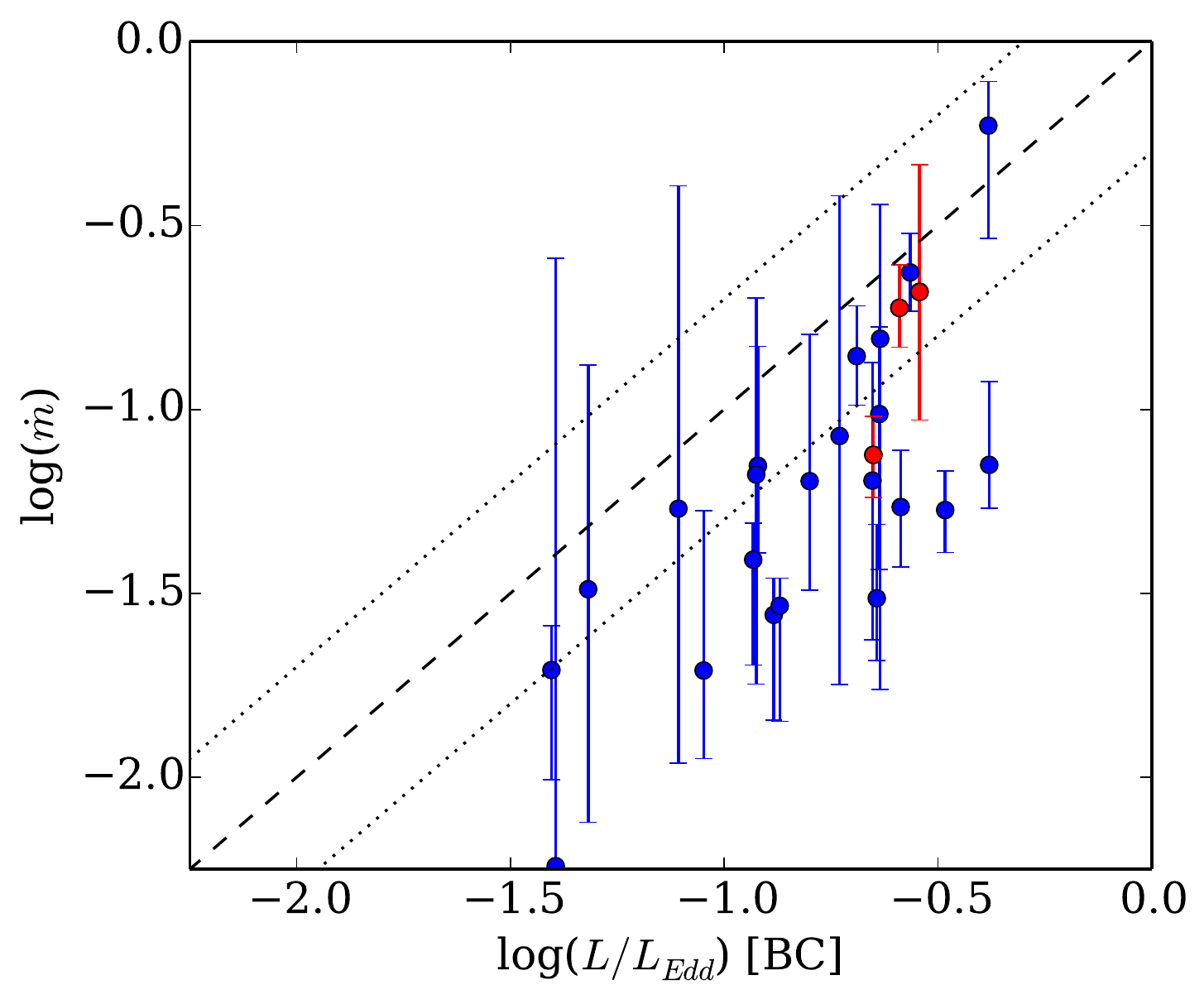}
 \caption{Same as Figure \ref{fig:mbh_comp}, but instead showing a comparison between
   $L/L_{Edd}$[BC], calculated directly from the observed spectra (using a bolometric correction [BC] factor; see Section \ref{sec:mbh_mdot}), and the median \mdot\ value from the Bayesian analysis, based on the median $a_*$, $M_{BH}$, and $\dot{M}$ values. The typical errors on $L/L_{Edd}$[BC] are at least as high as those on $M^{obs}_{BH}$, i.e. greater than 0.3$-$0.5 dex.
   }
 \label{fig:ledd_comp}
\end{figure}

As described in Section \ref{sec:sample}, we selected a sample of AGN, at roughly the same redshift, to evenly cover the widest possible range in $M_{BH}$ and $L/L_{Edd}$, as these two parameters, along with the spin, govern the physics of active SMBHs. Figures \ref{fig:mbh_comp} and \ref{fig:ledd_comp} compare our empirical measurements of $M_{BH}$ and $L/L_{Edd}$, as calculated from the \mgii\ emission line (see Section \ref{sec:mbh_mdot}) and a bolometric correction factor, to the parameters of the best-fit models from the Bayesian analysis for those AGN with a satisfactory fit.
Figure \ref{fig:mbh_comp} shows how the best-fit thin AD models have values of $M_{BH}$ that are within the error (0.3$-$0.5 dex) on the observed measurement of this parameter. In most cases, they are within the less conservative error estimate of 0.3 dex.

Figure \ref{fig:ledd_comp} compares our measured $L/L_{Edd}$ [BC] to the ``real'' $L/L_{Edd}$ ($\dot{m}$), which is obtained from the best-fit thin AD model to the observed SED of each AGN with a satisfactory fit.
This figure indicates that calculating $L/L_{Edd}$ using a bolometric correction factor tends to overestimate $L/L_{Edd}$. One of the main reasons for this discrepancy is that the use of standard bolometric correction factors does not take into account the wide range in possible spin parameters, which has a large effect on the SED \citep[see also][]{Netzer14}.

Figure \ref{fig:ledd_comp} also illustrates how the three AGN that could not be fit before an intrinsic reddening correction (the red points) are near the upper end of the $L/L_{Edd}$[BC] range.
AGNs with log $L/L_{Edd}$ $>$ $\sim$$-$0.5 might be powered by ``slim'' ADs, instead of thin ADs \citep{Netzer13}.
The SEDs of slim ADs can differ in several ways from the thin disk SED. It is possible that those SEDs that we could not fit without invoking reddening would be better fit by a slim AD SED.
Furthermore, three of the 5 AGN without satisfactory fits to the observed SEDs have log $L/L_{Edd}$[BC] of $-$0.558 to $-$0.408.
Therefore, these objects are potential candidates for slim ADs.

\section{Discussion}

\subsection{AGN Accretion Discs at z=1.55}

\subsubsection{$M_{BH}$ and $a_*$}
\label{sec:mbh_a}

\begin{figure*}
  \centering
  \includegraphics[width=140mm]{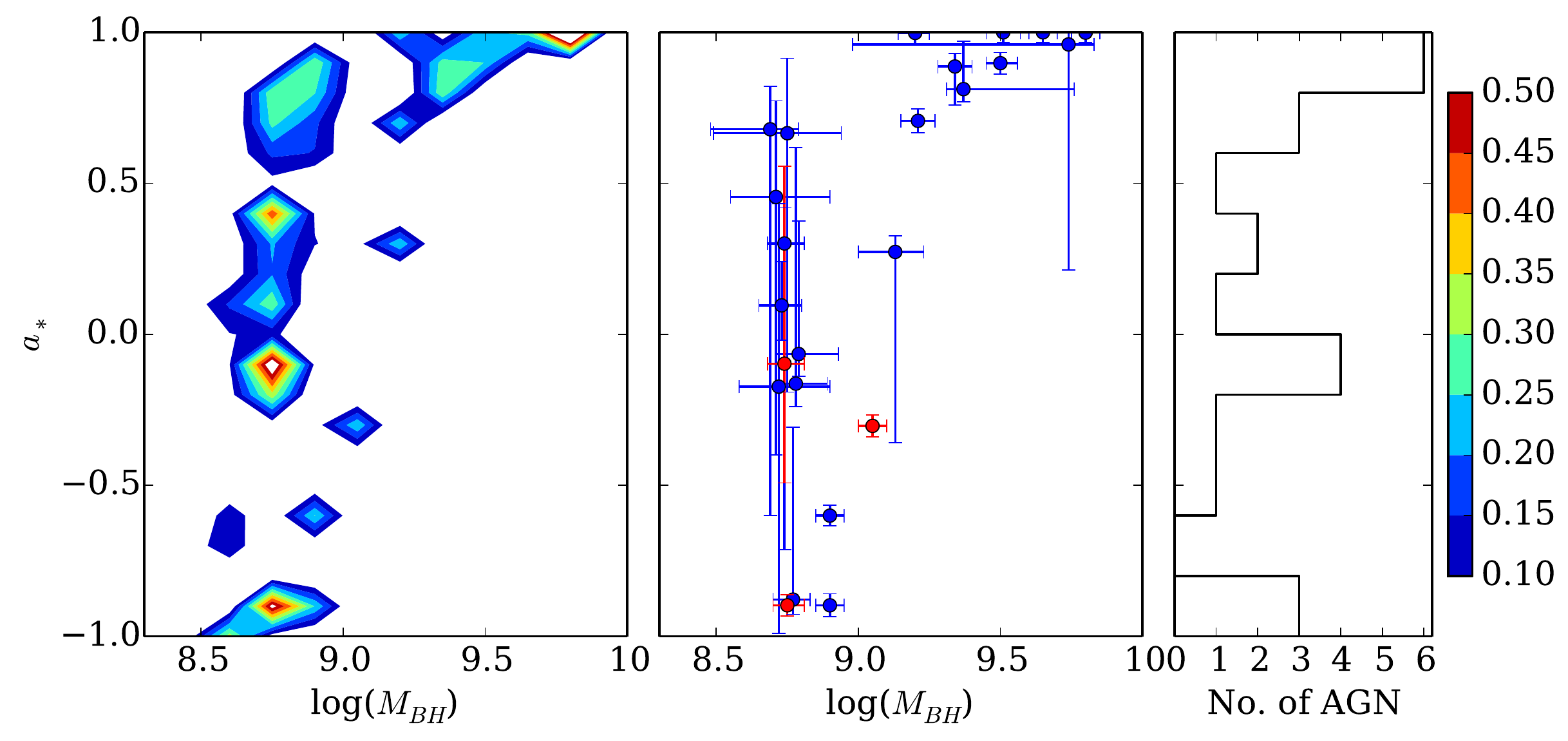}
  \caption{
    The spin parameter, $a_*$, as a function of $M_{BH}$. The left panel is a contour plot of the combined probability distributions in $a_*$ and $M_{BH}$ for the sources with satisfactory fits. The middle panel shows just the median $a_*$ and $M_{BH}$ values, where the red points are those sources requiring an instrinsic reddening correction. The right panel shows the distribution in the best-fit spin parameters.
  }
   \label{fig:mbh_a}
\end{figure*}

Current efforts to measure the spin of active SMBHs are limited to X-ray observations that can probe the innermost regions of the AD. Specifically, modeling the profile of the relativistic 6.4 keV K$\alpha$ line has been used to estimate the spin parameter in a small number of AGN with high-quality X-ray observations
\citep[][and references therein]{Fabian00,Brenneman13,Reynolds13,Risaliti13,Reis14}.
However, these measurements cannot distinguish between $a_* \approx 0$ and $a_*<0$ because the changes in the broad 6.4 keV line profile are very small. Moreoever, these measurements are currently limited to nearby AGN. A recent measurement of spin in an AGN at z $\sim$ 0.6 was possible because it is lensed \citep{Reis14}, but a way to measure the spin of active SMBHs at all redshifts is still needed.

In the current work, our Bayesian analysis determines which thin AD models have the highest probability of correctly explaining the shape of the observed SED and, therefore, constrains the value of the spin parameter for those cases where the thin AD model provides a satisfactory fit to the observed spectra (Table \ref{tab:master_table} and Figure \ref{fig:a_mbh_grid}).
In total, we find 25 such cases, including 3 where intrinsic reddening was taken into account.

We plot these values of $a_*$ versus $M_{BH}$ in Figure \ref{fig:mbh_a}. The diagram shows a large range of spin parameters, from very small ($-$1) to the largest possible value (0.998), with a trend for AGN with larger $M_{BH}$ to have larger spin values. We note however that our sample is not adequate for searching for this type of correlation since the objects were selected in such a way that must affect the BH mass and efficiency (and therefore spin) distribution.
Furthermore, on top of the formal uncertainties on $a_*$ that are marked in the diagram, there are other uncertainties that are related to the procedure we use to derive the spin. For example, for the 22 objects that had satisfactory fits with no correction for intrinsic reddening or a disc wind, it is possible that some of these AGN indeed have intrinsic reddening and/or a disc wind.
We postpone the detailed investigation of these scenarios to the completion of the
project, when we expect to have high quality spectra of the remaining 9 AGN in the
sample.

Another characteristic of the measured spin parameters is the larger uncertainty on $a_*$ for sources where this parameter is close to the middle of the allowed range.
This is not related in a simple way to the data quality, but rather to the gradual change of the disc SED over a relatively large range of spin. Thus, very high spin values and very low spin values are more robust. It is also important to note that the exact shape of the short wavelength, Lyman continuum SED is strongly dependent on the accretion rate which, in some of the sources, has a large uncertainty (see Fig. \ref{fig:a_mdot_grid}). Therefore, the small uncertainty on the spin does not rule out a large uncertainty on the SED at those shorter wavelengths. We will return to this point in a future paper that will include the analysis of the available {\it GALEX} data for our AGN sample.

Despite the above uncertainties, it is intriguing that, except for one source, none of the AGN with $M_{BH}$ greater than $10^{9}$ \Msun\ have a spin parameter less than 0. In general, these higher mass sources tend to have spin parameters greater than $\sim$0.7, which corresponds to $\eta$=0.1. For AGN with $M_{BH}$ less than $10^{9}$ \Msun, the spin values are generally below $\sim$0.7.
This is not a correlation between $\eta$ and BH mass, but rather a distinction between the properties of two different mass groups. This is in line with recent works that discuss spin parameters in AGN with very massive BHs. \citet{Trakhtenbrot14} estimate the spins of the largest known BHs at redshifts of $\sim$1.5-3.5. The assumption of thin ADs powering these sources leads to very high spin values, similar to the ones found here.
\citet{Netzer14} studied the distribution of $L/L_{Edd}$ in large SDSS sub-samples. They find that, for $M_{BH}$ $>$ 10$^9$ \Msun, only BHs with spin parameter close to the maximum allowed value can produce a strong enough ionizing continuum and broad emission lines that have large enough EWs to be detectable in such samples. \citet{Netzer14} further argue that AGNs with very large BH mass will drop from samples like SDSS unless their spin parameter is very high. The high-mass objects in our current sample may be part of this population.

Taking the results of Fig. \ref{fig:mbh_a} at face value, we can make a comparison to what is predicted by theoretical models of the evolution of SMBH spin in AGN. There are two primary scenarios to describe this evolution, ``spin-up'' and ``spin-down.'' The ``spin-down'' scenario postulates that a series of accretion episodes with random and isotropic orientations will cause SMBHs to ``spin-down'' to moderate spins near $a_*$ $\sim$ 0, regardless of the final mass of the SMBH \citep{King08,WangJM09,LiYR12,Dotti13}.
The distribution we find for $a_*$, which spans the entire range from $-$1 to 0.998, is inconsistent with this scenario. Instead, we find many SMBHs that are ``spun-up'' to $a_*$ $>$ 0.5, even among the lower mass sources. There is also a cluster of sources near $a_*$ of $-$1, indicating they were ``spun-up'' to a high spin parameter in a previous accretion episode(s).
SMBHs can ``spin up'' when the BH grows primarily via a single prolonged accretion episode, or in the case of the most massive BHs, when there is even a small amount of anisotropy in the orientation of the accretion episodes \citep{Dotti13,Volonteri13}. Our results therefore favor scenarios where there is some preferred orientation for the accreting material, whether the accretion occurs via a single prolonged episode or many episodes.

\subsubsection{Ionizing Continuum}

An additional check of the consistency of the AD SEDs with AGN observations can be obtained by studying the predicted ionizing continuum and comparing it with observations of several strong emission lines. The relative intensity of the lines, and their equivalent widths,
are related to the ionization parameter in the broad emission line region (BELR), the mean energy of the ionizing photons, and the covering factor by high density gas near the BH \cite[e.g.][and references therein]{Netzer13}.
For example, the relatively high intensity of the strong \civ\ line is usually an indication of both a high ionization parameter and a relatively hard ionizing continuum.

We calculated the mean energy of an ionizing photon for all of our best fitted AD models. These numbers are given in Table \ref{tab:master_table}.
The numbers depend on  \mdot\ and $a_*$ and range from 1.26 Ryd (J0850+0022; \mdot=0.054, $a_*$$=$$-$0.897) to 2.45 Ryd (J1152+0702; \mdot=0.59, $a_*$$=$0.998).
For a comparison, the mean energy of an ionizing photon in a $L_{\nu} \propto \nu^{-1.5}$ SED, extending from 13.6 to 200 eV, is 2.31 Ryd.
Such slopes have been estimated in a large number of AGN by connecting the observed point at 1000$-$1200\AA\ with the observed X-ray continuum below 1 keV \cite[][and references therein]{Davis11,Shull12,Stevans14}.
We also calculated the expected equivalent width (EW) of the Ly$\alpha$ line assuming a covering fraction of 10\% for the BEL gas and case B recombination conditions.
The numbers range from 18$-$120 \AA\ and are in general agreement with the observed EWs.
As noted above, a relatively small change in \mdot, within the uncertainties allowed here, can cause a significant change in these values.
We will discuss these issues in detail in the next paper in this series.

\subsection{Reddening in Host Galaxies of AGN}

If the thin AD model adopted in this work does indeed explain the emitted SED of the AGN in our sample, and if we assume no wind, then the host galaxies of approximately 1/3 of the AGN in our sample contain enough dust to significantly affect the observed SED (intrinsic reddening). Of course, as shown in Section \ref{res:winds}, some of the observed SEDs can be fit without any intrinsic reddening correction, but by adding a disc wind to the thin AD model. Because of the uncertainties in the models and in the input parameters to the model, it is not possible to determine whether accounting for intrinsic reddening or adding a disc wind provides the better solution for each source.
Furthermore, it is entirely possible that the observed SED could be affected by both intrinsic reddening and a disc wind.

However, our results do indicate that there are four AGN that are best fit by the thin AD model only after correcting the observed SED with the Milky Way extinction curve. This is because the model initially overestimates the luminosity at the continuum region around 2200 \AA. Other extinction curves, such as the SMC curve, clearly provide poorer fits. Several earlier studies claimed that this 2175\AA\ feature is not observed in AGN spectra. This leads to speculations about the nature of the dust grains in AGN host galaxies, in particular the lack of small graphite-type grains that are thought to be the main contributors to the absorption around this wavelength \citep{Maiolino01,Hopkins04}. Our observations suggest that much of the earlier speculations may simply be the result of inaccurate, limited waveband observations.
The apparent presence of this bump in 4 out of the 30 AGN in our sample indicates that the reddening in at least some small percent of AGN host galaxies can be best described by the Milky Way extinction curve. Furthermore, we do not find any cases where the SMC extinction curve allows for a better SED fit than the simple power-law or Milky Way extinction curves, contrary to previous claims  \citep[e.g.][]{Hopkins04,Glikman12}.

\subsection{Unusual SEDs}

Several AGN in our sample have unusual observed SEDs. In particular, J1108+0141 has an unusual and strong small blue bump (the spectral region around $\sim$2200$-$3900 \AA). While the model with the highest posterior probability for J1108+0141 has a $\chi^2$ value within our threshold for a satisfactory fit, the shape of the thin AD SED clearly does not match the observed SED.
The observed spectrum for J0148+0003 has a very different overall shape from the rest of the sample. The spectrum curves downwards strongly towards shorter wavelengths. Correcting for intrinsic reddening, using an $A_V$ of 0.45, gives the spectrum a similar shape to other AGN in the sample. We are then able to find a marginal thin AD fit to the spectrum. We note that this amount of reddening is at the end of the $A_V$ distribution found here and clearly does not represent type-I AGN. This object may belong to the population of extremely red AGN studied, e.g. by \citet{Richards03}, \citet{Glikman12}, and \citet{Banerji12,Banerji13}. \citet{Glikman12} find sources with reddening up to $E(B-V)\sim$ 1.5.

Finally, our fitted AD models do not take into account the possibility of slim ADs that may be more relevant to sources with $L/L_{Edd} >$0.2. Such SEDs are more difficult to calculate, and present day models are far more uncertain than those used here. We defer this kind of discussion until we obtain a larger sample with high quality spectra of such sources.

\section{Conclusions}

This work is the first in a series of papers with the aim of testing how the three main parameters that govern the physics of active BHs $-$ mass, spin, and accretion rate $-$ determine the observable attributes of AGN.
Using a unique sample of 30 AGN in a narrow redshift range around z $\sim$ 1.55, covering a range of $\sim$0.04 to 0.7 in $L/L_{Edd}$[BC] and $\sim$$2 \times 10^{8}$ to $4 \times 10^{9}$ M$_{\sun}$ in \mbh, and observed with the {\it X-shooter} instrument at the VLT, we fit thin AD models to observed SEDs.
We use a Bayesian method to consider models with varying $M_{BH}$, $\dot{M}$, inclination, and spin parameter ($a_*$).
With this method, we are able to fit 22/30 of the SEDs (Section \ref{res:fitting}). Of the remaining 8 AGN, we are able to find satisfactory fits to 3 SEDs and marginal fits to 4 SEDs, after correcting for intrinsic reddening (Section \ref{res:red}). Some SEDs were best fit when using the Milky Way extinction curve to correct the spectrum. Alternatively, some of these 8 sources could be fit by adding a disc wind to the model, instead of correcting the observed SED for instrinsic reddening (Section \ref{res:winds}). These results are in constrast to much of the earlier work that could not fit thin AD models to observed AGN SEDs, most likely because this earlier work was hindered by possible variability and/or a limited observed wavelength range. The results of the current work indicate that thin ADs are indeed the main power houses of AGN.

Based on the satisfactory fits to the observed SEDs, we find a wide distribution in the spin parameter, $a_*$, covering the entire range from $a_*$ = $-$1 to 0.998. This range in $a_*$, along with the concentration of the most massive BHs at the highest spin parameters, is consistent with the ``spin-up'' scenario of BH accretion, rather than the ``spin-down'' model. Our results also indicate that $\dot{m}$, in general, is smaller than the values obtained by using simple bolometric correction factors ($L/L_{Edd}$[BC]).

The next paper in the series (Mej\'{i}a-Restrepo et al. in preparation) will study in detail the emission lines in all the SEDs in our sample and look for trends in the profile shapes, strengths, and other properties, with BH mass, spin, and accretion rate. Finally, the analysis of the major emission lines $-$ \civ, \mgii, \Hbeta, and \Halpha\ $-$ will allow us to directly compare \mbh\ measurements that are based on their profiles.

\section*{Acknowledgments}

We thank Jian-Min Wang for useful discussion. We thank George Becker, Andrea Modigliani, and Sabine Moehler for their advice and assistance with the data reductions.
We thank an anonymous referee for helpful comments on the manuscript.
We thank the DFG for support via German Israeli Cooperation grant STE1869/1-1.GE625/15-1.
Funding for this work has also been provided by the Israel Science Foundation grant number 284/13.
PL received support from Fondecyt Project 1120328 and the Center of Excellence in Astrophysics and Associated Technologies (PFB 06).

\bibliographystyle{mn2e}

\bibliography{full_bibliography_tau}

\appendix

\section{bayesian analysis}
\label{app:bayes}

In general terms, the Bayesian analysis is based on the derivation of
the posterior probability $P(H|D,I)$ for a hypothesis $H$, given the
available data $D$, and any {\bf prior} information $I$ about $H$,
known before the analysis of the current data. The posterior
probability can be expressed as \citep[e.g.,][]{Sivia10}:

\[ P(H|D,I) = \frac{P(D|H,I)\times P(H,I)}{P(D|I)} \]

In our case the hypothesis is that a particular model $m = m(M^{mod},
\dot{M}^{mod}, a, \theta)$ is a good representation of the spectral
energy distribution seen in the observed X-shooter data, which are
given by $D_{i} \pm \sigma_{D_{i}}$, with $i$ the number of
independent data measurements.

Notice that to determine whether model $m_1$ is better than model $m_2$, we only need to find:

\[ \frac{P(m_1|D,I)}{P(m_2|D,I)} = \frac{P(D|m_1,I) \times P(m_1|I)}{P(D|m_2,I) \times P(m_2|I)} \]

As usual, the probability of observing the measured data $D$ if the model $m_k$ was true can be computed as:
\[ P(D|m_k,I) \propto \exp(-\sum_i (m_i - D_i)^2/2\sigma_{D_i}^2) = \mathcal{L}(m_k) , \] 
where $\mathcal{L}(m_k)$ (which can be recognised as $\propto \exp
(-\chi^2/2)$) is refered to as the likelihood of model $m_k$.

Our prior information $I$ corresponds to the observed values for the
black hole mass and accretion rate, $M^{obs}$ and $\dot{M}^{obs}$, and
their uncertainties $\sigma_{M}$ and $\sigma_{\dot{M}}$.
We have no prior information on $a_*$ and $\theta$, so each value of $a_*$ and 
$\theta$ that we consider has equal probability.
The derivations of $M^{obs}$ and $\dot{M}^{obs}$ are given by:
\begin{eqnarray*}
  M^{obs} = A_{1}\times {\rm FWHM({\rm Mg}II)}^{2} \times L_{1}^{\alpha} \\
  \dot{M}^{obs} = A_{2} \times L_{2}^{3/2}/M^{obs}
\end{eqnarray*}
where $A_1$ and $A_2$ are scaling factors, $L_1$ and $L_2$ are
luminosities derived from two different continuum ranges, and $\alpha$
is the index of the radius-luminosity relation. 

Then:
\begin{eqnarray*}
  P(m_k|D, I) = P(m_k|D,M^{obs},\dot{M}^{obs}) \propto \\
  \propto \mathcal{L}(m_k) \times P(m_k|M^{obs},\dot{M}^{obs})
\end{eqnarray*}

Using the fact that $\dot{M}^{obs}$ and $M^{obs}$ have no constraints
on parameters $a_*$ and $\theta$ \footnote{This corresponds to the marginalization of $M^{mod}$ and $\dot{M}^{mod}$ from the suit of model parameters, assuming delta function probability distributions for parameters $a_*$ and $\theta$.}, and applying the product rule of
probabilities, we can write:
\begin{eqnarray*}
  P(m_k|M^{obs},\dot{M}^{obs}) = P(M^{mod},\dot{M}^{mod}|M^{obs},\dot{M}^{obs}) = \\
   = P(M^{mod}|M^{obs},\dot{M}^{obs}) \times P(\dot{M}^{mod}|M^{mod},M^{obs},\dot{M}^{obs})
\end{eqnarray*}

Assuming a Gaussian probability distribution for $M^{obs}$  with standard deviations equal to $\sigma_{M}$ 
and the fact that $\dot{M}^{obs}$
has no constraints on $M^{mod}$, the first term in the previous
expression can be written as:
\begin{eqnarray*}
  P(M^{mod}|M^{obs},\dot{M}^{obs}) = P(M^{mod}|M^{obs}) \propto \\
    \propto \exp(-(M^{mod}-M^{obs})^{2}/2\sigma_{M}^{2})
\end{eqnarray*}

For the second term we need to determine the probability of $\dot{M}^{mod}$ given $\dot{M}^{obs}$,  $M^{obs}$ {\bf and} $M^{mod}$.
Since $M^{mod}$ completely determines the mass value\footnote{This corresponds to the marginalization of $\dot{M}^{mod}$ assuming a delta function probability distribution for parameter $M^{mod}$ and Gaussian probability distributions for $M^{obs}$ and $\dot{M}^{obs}$: 
$\int \delta(M-M^{mod}) \exp(-(M^{obs}-M)^{2}/2\sigma_{M}^{2}) \exp(-(\dot{M}^{obs}\!-\dot{M}^{mod})/2\sigma_{\dot{M}}^{2})\ dM$}:
\begin{eqnarray*}
  P(\dot{M}^{mod}|M^{mod},\dot{M}^{obs},M^{obs}) \propto \\
    \propto \exp(-(\dot{M}^{obs}|_{M^{mod}}-\dot{M}^{mod})^{2})/2\sigma_{\dot{M}}^{2})
\end{eqnarray*}

Since $\dot{M}^{obs} \propto 1/M^{obs}$, finally:  
\begin{eqnarray*}
  P(H|D,I) = \exp (-\!\sum_{i} \frac{(D_{i}-m_{i})^{2}}{2\sigma_{D_{i}}^{2}}) \times \\
    \times \exp(-(M^{obs}\!-M^{mod})^{2}/2\sigma_{M}^{2}) \times \\
    \times \exp(-(\dot{M}^{obs}\! \times\! \frac{M^{obs}}{M^{mod}}-\dot{M}^{mod})^{2}/2\sigma_{\dot{M}}^{2})
\end{eqnarray*}

\bsp

\label{lastpage}

\end{document}